\documentclass[english,a4paper,11pt]{article}
\usepackage[T1]{fontenc}
\usepackage[ansinew]{inputenc}
\usepackage{babel}
\pdfoutput=1

\usepackage[left=3cm,top=3cm,bottom=3cm,right=3cm]{geometry}
\usepackage{graphicx,amsmath,booktabs,color,times,mathptmx}
\usepackage{natbib}
	
	\bibpunct{(}{)}{;}{a}{}{,}	
	
\let\oldtabular\tabular 
\renewcommand{\tabular}{\small\oldtabular}

\begin{document}
\author{\normalsize Giangiacomo Bravo\thanks{Department of Social Sciences, University of Torino and Collegio Carlo Alberto. 
Email \mbox{giangiacomo.bravo@unito.it}}}
\title{Cultural commons and cultural evolution}
\date{}
\maketitle

\begin{abstract}
Culture evolves following a process that is akin to biological evolution, although with some significant differences. At the same time culture has often a  collective good value for human groups. This paper studies culture in an evolutionary perspective, with a focus on the implications of group definition for the coexistence of different cultures. A model of cultural evolution is presented where agents interacts in an artificial environment. The belonging to a specific memetic group is a major factor allowing agents to exploit different environmental niches with, as a result, the coexistence of different cultures in the same environment. 
\end{abstract}

\section{Introduction}

Richard Dawkins, who first introduced the term \emph{meme}, described it as a replicator: a unit of cultural transmission which propagate itself in the meme pool in analogy to what genes do in the gene pool \citep[Chap. 11]{Dawkins1976}. More generally, he proposed a deep analogy between biological and cultural evolution, with the latter encompassing memes that replicate trough imitation, mutate when new cultural variants are produced and compete for human brain resources. While this analogy must not be taken too far --- critics underline, for instance,  that memes have often weakly-defined boundaries, replicate with much less fidelity than genes or that selection pressures on memes derive from little understood mechanisms \citep[e.g.][]{Atran2001,RB2005,Wimsatt1999} --- at a more general level there are little doubts that some kind of evolutionary process underlies culture development and change \citep{Boyd1985,Dawkins1976,Mesoudi2004,Plotkin1993,Rogers2008,RB2005}.

Thinking to culture as an evolutionary process is not without problems. Among them, especially intriguing is the fact that, given a constant or sufficiently slow changing selective pressure, replicators should converge in the long run into a small number of highly adapted forms. However, as an enormous number of genera and species live side to side in the biological world, cultural variants coexists in the cultural world. The  paradox of the presence of a large biodiversity, resulting from apparently converging selective pressures, is resolved in natural sciences using the concept of niche adaptation and taking into account the constant moving of species between regions \citep{Hubbell2001,Sugihara2003,Tilman2004}. However, few social scientists have taken the problem from this point of view, focusing instead their research on mechanisms connected to some structural characteristics of human societies rather than with the cultural environment itself \citep[e.g.][]{Axelrod1997a}.

We present here a model of cultural evolution where agents are able to exploit different niches on an artificial environment. This ability is strictly linked with their belonging to a memetic group (that can be viewed as a culture). More precisely, agents belonging to a group share a part of their ``memotype'' \citep{Speel1997}. Instead of imitating any other agent in the population, agents updating their strategies tend follow the ones already present in their own group. However, agents can also choose to change their group: a fact that produces a  multilevel selection process. Note that, while multilevel selection is a source of unending debate in biology \citep[for a recent review see][]{Kohn2008}, it may be more plausible in reference to cultural evolution \citep{Retal2003,RB2005,Soltis1995,WS1994}. In our model, belonging to a specific memetic group allows agents to exploit different niches of their environment with, as a result, the coexistence of different memotypes in the same population. This in contrast with a model where selection takes place only at the level of individual agent leading instead to the fixation of a single memotype. Finally, we show that a situation when changes in the meme pool of agents also modify the environment where they interact favor the establishment of coexisting memetic groups in the system and, more generally, it is able to model realistic evolutionary processes.

The remaining of the paper is organized as follows. Section \ref{theory} illustrates the research background. Section \ref{model} defines our cultural evolution model and presents the corresponding results. Section \ref{coevolution} defines and presents the results of the niche construction model, where agents' evolution affects also the state of the environment. Finally,  Section \ref{discussion} discusses the results.

\section{Groups, niches and cultural differences}\label{theory}

One of the seminal and most well known researches trying to model culture evolution and dissemination is Robert \citet{Axelrod1997a} work on \emph{Social influence}. Its main result is that  cultural differences can be sustained, despite the effect of mechanisms like homophily and social influence, by the creation of boundaries between ``cultural regions''. In Axelrod's model, $n$ agents interact on a regular lattice, typically a 2D, $10 \times 10$ grid. Each agent has a cultural state defined by a vector of $F$ elements (cultural features), which can assume any of $q$ different discrete values. Agents' initial state is defined randomly. Subsequently, a pair of neighboring agents is selected in each time step, with their interaction probability depending on the respective states: it is zero when the couple shares no features and increases proportionally to the number of shared features (homophily assumption). When two agents interacts, one of them copies one of the unshared features of the other (social influence). The simulation continues until a stable state is reached, either because all agents reach the same state or because stable boundaries emerge between cultural regions composed by agents sharing no features. Axelrod shows that, under a wide range of condition, the latter outcome is more likely: a fact that tend to maintain over time cultural differences in the system. 

This result is interesting, especially because it holds despite the fact that the interactions occur in a regular space presenting no possibilities of niche differentiation and that the modeled features possess no adaptive meaning for the agents. The Social influence model  represents hence an alternative to biological explanations of biodiversity based on the niche adaptation concept. Unfortunately, subsequent works have shown that Axelrod's results depend critically on a few implausible assumptions, including the absence of noise in the transmission process \citep{Klemm2003a} and discrete cultural variants \citep{Flache2006}, possibly drawn from a large set \citep{Klemm2003}. Moreover, Axelrod's results hold only under the assumption of \emph{perfect homophily}, i.e. a null interaction probability between agents sharing no cultural features. The creation of impermeable  boundaries represents indeed the key factor able to guarantee pluralism \citep{Axelrod1997a}. The problem with this assumption is that no boundary is perfectly impenetrable in the real world, while relaxing it leads inevitably to an homogeneous state in the long run.\footnote{\citet{Flache2006} first discussed this fragility of the model, even if they apparently failed to recognize its full power in undermining Axelrod's conclusions.}

The example in Figure \ref{fig:epsilon} illustrates the effect of introducing an arbitrarily small probability of interaction between agents sharing no features ($\varepsilon$ ) in a system where six agents interact on a one-dimensional lattice.\footnote{A 1D system is used here for simplicity, but the same argument holds for multidimensional systems as well, including the regular 2D lattice used by Axelrod.} Each agent has five cultural features that can assume any integer value between zero and nine. Line (a) presents an equilibrium  situation as described by Axelrod, with a stable cultural boundary between the third and the fourth agent. Line (b) shows the same situation but for a small $\varepsilon$  probability of interaction between the two middle agents sharing no cultural trait. The probability that the fourth agent is selected and copies one feature from the third one is $\varepsilon / 2n$. Taking into account that also the third agent has the same probability of copying one trait from the fourth one, the overall interaction probability is $\varepsilon / n$. No matter how small, this probability is strictly greater than zero and, given a sufficient amount of time, the fourth agent will actually copy one feature from the third one (or vice versa). This leads to the situation depicted in line (c), with an increase of the interaction probability between the fourth agent and \emph{both} its neighbors. More generally, the system moves away from the stable state in a process that is self-reinforcing. Actually it is quite intuitive that a system with a probability of interaction between agents sharing no features is greater than zero possesses only one stable state: the one in which all agents are identical. Simulations based on the Axelrod's model confirm this intuition.

\begin{figure}[t]%
\centering
\fbox{\parbox{\linewidth}{\centering
\begin{tabular}{lccccccccccc}
(a) & 01234 & (0) & 01234 & (0) & 01234 & (0) & 56789 & (0) & 56789 & (0) & 56789 \\
&&&&&&&&&&&\\
(b) & 01234 & (0) & 01234 & (0) & 01234 & ($\varepsilon / n$) & 56789 & (0) & 56789 & (0) & 56789 \\
&&&&&&&&&&&\\
(c) & 01234 & (0) & 01234 & (0) & 01234 & ($0.2/n$) & 06789 & ($0.8/n$) & 56789 & (0) & 56789 \\
\end{tabular}
}}
\caption{Effect of the introduction of an arbitrarily small probability of interaction between agents with no traits in common in a 1D version of Axelrod's Social influence model. The probability of interaction between two neighbors is in parenthesis.}
\label{fig:epsilon}
\end{figure}

Our argument is that Axelrod's model most fundamental problem derives from the uniform space where agent interacts, which does not allow the exploitation of different niches by different cultures. This is vastly different from what happens for biological evolution, and also from what we know from the functioning of social systems, where different groups are usually able to find vastly different ways of making a living by exploiting different resources. For instance, pastoral communities can live close to farmers by using different natural resources. Moreover, higher level specialists, like merchants, can rely only indirectly on natural resources by exploiting niches built up by the work of other individuals.

The link between different environments (or niches) and different cultures is supported by some classical anthropological studies \citep{Harris1979}. \citet{Henrich1998} developed a model where different subpopulations are able to evolve specific adaptations to the different environments where they live, despite a significant amount of migration among them. The core of the model is ``conformist transmission'', i.e. the  fact that agents tend to imitate disproportionately the most commons behavior in their group. From our point of view, it is especially important to remark that conformist transmission, while it favors within-group cultural homogeneity, appears to be able to maintain between-group differences. However, the model achieves this result by building  separate environments with opposite adaptive needs. It does hence not explain how cultural diversity can be maintained when agents interact \emph{in the same environment}. On the contrary, the effect of conformist transmission in a single environment setting is  likely to lead to a homogeneous state. 

An alternative approach is thinking about cultures in terms of biological species \citep{Mace2005, PM2004}. The core of the species concept is that organisms from different species cannot interbreed successfully. Biodiversity derives hence from niche adaptation, but it is maintained thanks to strong barriers to the gene flow. If the analogy between cultures and species holds, it implies that the ``interbreeding'' among cultures must be limited in order to maintain diversity. This is actually what happens in human groups, where a number of mechanisms exist to limit cultural and possibly genetic influences from other groups \citep{PM2004,RB2001}. While this may surprise cosmopolitan citizens of modern cities, empirical studies actually show that a large proportion of cultural transmission occurs phylogenetically within groups, including (of course) families and (more generally) homogeneous cultural groups \citep{Guglielmino1995,Mace2005}. 

The consequence of a reduction the meme flow among cultures is a decrease of within-group variation and a corresponding increase of between-group variation: a situation that increases the possibility that selection processes occur also at the group level, as acknowledged also by group-selection skeptics \citep[e.g.][Chap. 13]{Dawkins1989}. This is actually one of the reasons why group selection processes are more likely to be a significant force in cultural than in biological evolution \citep{Retal2003,RB2005,Soltis1995,WS1994}. Another reason is that, although appealing, the cultures-as-species metaphor is wrong in one important detail: unlike biological organisms, that are rarely able to acquire DNA from other living beings,\footnote{This happens mainly in bacteria and other microbes: asexual organisms for whom species are weakly defined.} memes in human brains are relatively easy to change at any point of an individual's life. This has important consequences,  because intergroup migration represents a major factor undermining group-selection processes in biological evolution \citep{MaynardSmith1976}. On the contrary, human migrants often acquire the culture of their new groups, making hence migration less effective in reducing between-group variation. This, along with the high levels of intergroup conflict typical of human societies \citep[e.g.][]{Soltis1995}, makes group selection processes plausible, if not probable, for cultural evolution \citep[Chap. 6]{RB2005}: a fact that justifies our choice of modeling a multilevel selection process (see below).

Before proceeding with the model definition, one more point should be carefully discussed: the value of a given culture for the people who share it. Using a strict ``meme-eye view'' the only interest of the memes would be to create as many copies of themselves as possible. However, from the point of view of individuals hosting these memes, their value is clearly linked to the opportunities that they offer in finding adequate ways of making a living. Similarly, the adaptiveness of a given group to its environment depends on the culture shared by its members. Cultures allowing groups to exploit some environmental niche hence possess the value of a shared resource. While natural shared resources often present ``tragedy of the commons'' problems leading to their depletion or destruction \citep{Hardin1968}, cultural commons tend instead to increase their value with use \citep{Boiller2007,Hess2007}. For instance, the value of a software or of a piece of scientific knowledge increases as more people use it. The problem here is indeed not overuse, like in natural commons, but reaching the critical mass that makes a given culture self-sustaining. 

In order to coexist, cultures should also allow their members to exploit different niches of the social or of the natural environment (or both). Otherwise, in absence of perfectly impermeable boundaries, selection processes tend to lead to the fixation of a single culture. Each group must hence be able to define and use its own ``cultural commons'', i.e. the shared cultural resource allowing the exploitation of a specific niche in its environment. Note that, even if cultural commons do not suffer from overuse, still they have to be maintained and protected from ``erosion'', i.e. the undermining of their internal coherence following uncontrolled changes. However, no sustainable use of a commons is likely without the definition of boundaries, including limits of its users' group \citep{Ostrom1990}. This point parallels hence the above group-selection argument in supporting a strong influence of group barriers in making the coexistence of different cultures possible.

Our arguments can be summarized in two hypotheses. First any coexistence among cultures need boundaries. They may not be perfectly impermeable, as in Axelrod's model, but still they should be strong enough to permit the definition of in-group and out-group members. Most of the times, boundaries in human societies are quite strong and produce within-group cooperation and favoritism and between-group competition or conflict \citep{Bernhard-etal2006,Iida2007,RB2001,Soltis1995,Tajfel1986}. Nevertheless, boundaries are not a sufficient condition since, in absence of the possibility of exploiting different niches, competition will lead to the disappearance of the weakest groups and, in the long run, to the fixation of a single culture. Our second hypothesis is hence that cultures can coexist by allowing their members to exploit different niches of their social-ecological system,\footnote{For a definition of social-ecological systems, see \citet{Berkes1998}.} just as specialization (and often speciation) allows different organisms to exploit different ecosystem niches. The models presented in the next sections will explore these ideas.

\section{The cultural evolution models}\label{model}

In order to test our hypotheses, we designed two simulation models. The first one, named \emph{Base}, represents our benchmark. It defines our artificial agents and environment, but supports no group processes or boundaries. In the \emph{Group} model, boundaries among groups of agents are instead explicitly defined, even if they are not perfectly impermeable. 

\subsection{\emph{Base} model definition}

\begin{figure}[t]
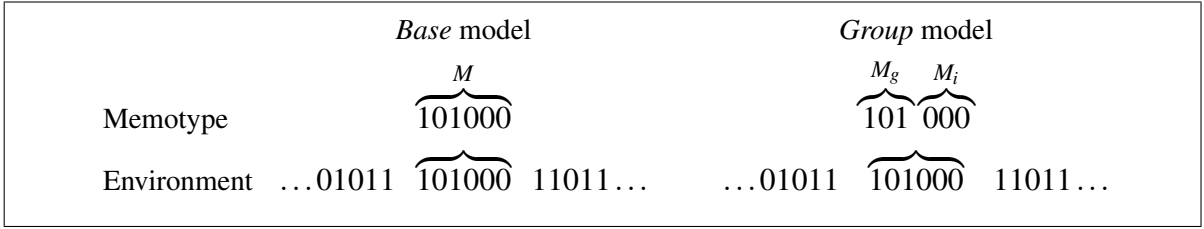
%
\fbox{
\parbox{\linewidth}{\large
\[
\begin{array}{lc@{}c@{}c@{\hspace{1cm}}c@{}c@{}c}
& & \text{\normalsize \emph{Base} model} & & & \text{\normalsize \emph{Group} model} & \\
\noalign{\vskip3pt}
\text{\normalsize Memotype} & & \overbrace{101000}^{M} & & & \overbrace{101}^{M_g}\overbrace{000}^{M_i} &\\
\text{\normalsize Environment} & \ldots 01011 & \overbrace{101000}^{} & 11011 \ldots & \ldots 01011 & \overbrace{101000}^{} & 11011 \ldots
\end{array}
\]
}}
\caption{Example of memotype-environment correspondence in the \emph{Base} and the \emph{Group} models.}%
\label{fig:example}%
\end{figure}

The base model is formed by $N$ agents, representing organisms, interacting on an abstract environment.\footnote{All models have been implemented in C++. Codes are available upon request.} The environment $E$ is simply a sequence  of $n$ binary numbers. It is randomly determined at the beginning of each run and remains subsequently constant. Each of the $N$ agents possess a  ``memotype'' $M$ formed by a sequence of $m = 6$ binary numbers. This gives a total of 64 possible memotypes. Adopting a population of 256 agents, we have an average of 4 agents per memotype. At the beginning of each run, memotypes are determined randomly, while they are subsequently free to evolve. During each period of the game agents earn a variable number of points. The payoff of each agent depends on the correspondence between its memotype and the environment. More specifically, each time that $M$ corresponds exactly to a sub-sequence of $E$ the agent earns one point (Fig. \ref{fig:example} left). The period payoff is simply the sum of all points earned by a given agent in a single period.

A fundamental feature of the model is the imitation process occurring at the end of each period. During imitation, one of the agents with the lowest period payoff copies the memotype of one of the agents with the highest period payoff.  However, a small mutation probability $\mu$ exists. When a mutation occurs the new memotype of the low payoff earner is randomly defined. Note that the term mutation should not be taken here in a strict biological meaning. Rather, it stands for innovations and new ideas entering the game and allowing for the exploration of new parts of the environment. 

Before proceeding with the definition of the \emph{Group} model, it is important to understand how the length of the environment  influences the number of niches where agents can make their living. We define a niche as a sub-sequence of $E$ of length $m$ leading to the maximal period payoff for an agent holding the corresponding memotype. For instance, let us suppose the case where $E = \{0,0,0,1,1,1\}$ and $m=2$. Both the memotypes $M_{00}=\{0,0\}$ and $M_{11}=\{1,1\}$ earn two points per period, while  $M_{01}=\{0,1\}$ earns one point and $M_{10}=\{1,0\}$ zero points. The two memotypes $M_{00}$ and $M_{11}$represent cultural variations that are neutral in respect to the selection process, both leading to the maximal payoff. Following the definition above,  $E$  possesses hence two niches.

\begin{figure}[t]%
\centering
\includegraphics[width=0.45\textwidth]{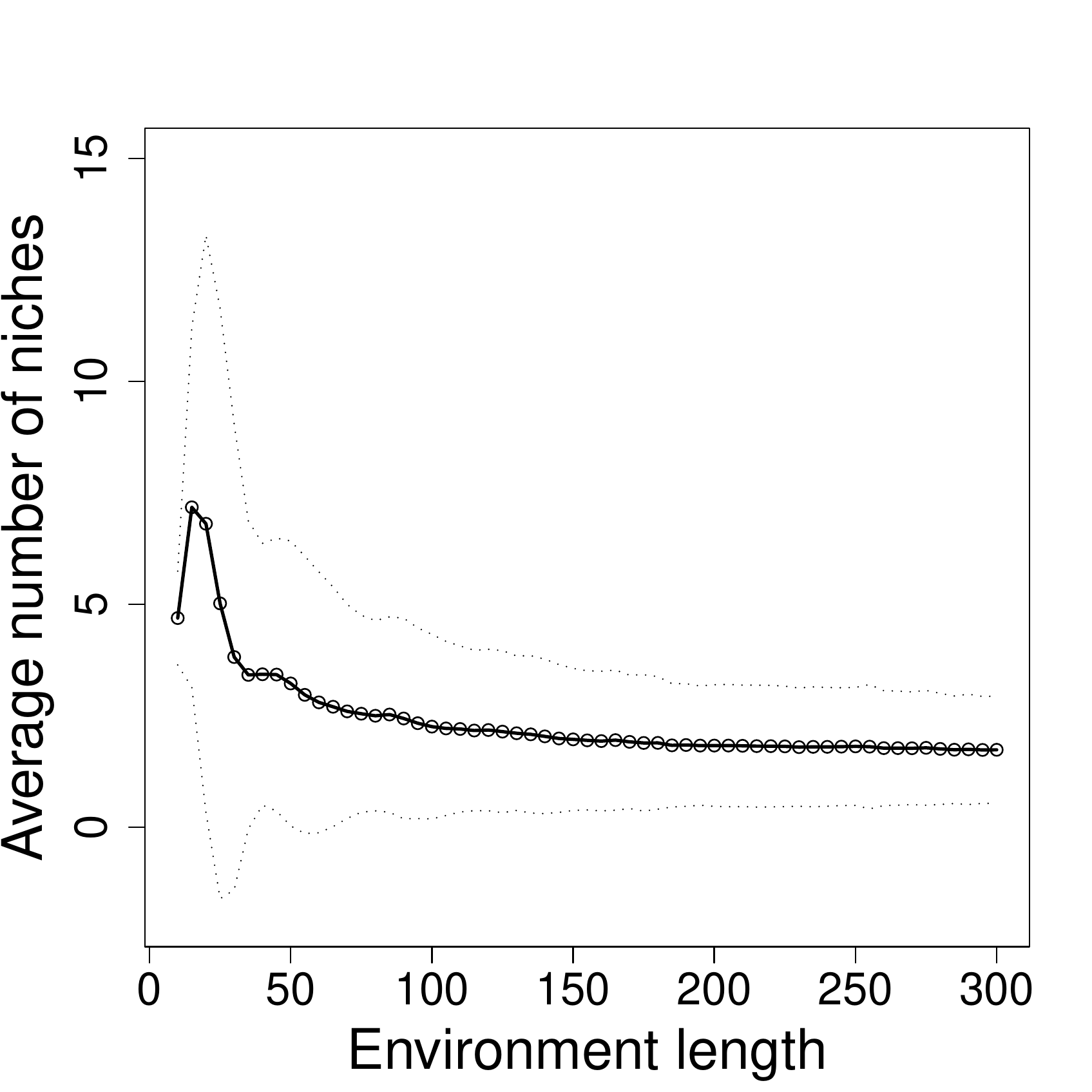}%
\caption{Average number of niches per environment lengt for a memotype of length $m=6$. Standard deviations are represented as dotted lines.}%
\label{fig:environment}%
\end{figure}

In order to choose the environment length to use in the model, we explored the number of niches in our environment for $m=6$ and $n \in \{10,15,\ldots,300\}$. The results of a sample of 10000 environments for each value of $n$ are reported in Figure \ref{fig:environment}. It is clear that the number of niches (but also the standard deviation) reach a peak at $n \approx 3m$, reduces rapidly up to $n \approx 100$, while for higher values of $n$ it shows a continuous, although extremely slow, decline .\footnote{Note that a similar trend of peak as subsequently slow decline is present also for other values of $m$.} 

It would be tempting to choose for our simulations values of $n$ leading to a high number of niches. However, the standard deviation corresponding to low values of $n$ is extremely high (Fig. \ref{fig:environment}), a factor that could undermine the stability of our results. We hence preferred to use the smaller value of $n$ leading reasonably close to the level of 2 niches. This resulted to be $n=150$, corresponding to an average of 1.97 niches per environment.

\subsection{\emph{Group} model definition}

The \emph{Group} model introduces  boundaries in the game, making some memes easier to adopt than others. More specifically, the \emph{Group} model maintains all the elements of the \emph{Base} one except for the fact that the memotype of each agent is now formed by the concatenation of two sequences, $M_g$ and $M_i$, composed by $m_g$ and $(m - m_g)$ binary numbers respectively. The former identifies the section of the memotype common to each group while the latter defines the agent's specific memotype. Once the two sequences are concatenated, the  memotype-environment correspondence is checked as in the \emph{Base} model (Fig. \ref{fig:example} right). The imitation process occurs as in the previous case, but it now concerns only $M_i$. Moreover, imitators and imitation targets are selected inside each group. This means that, for each $M_g$ sequence, one hight-earner agent and one low-earner agent are selected, with the latter copying the former's $M_i$. The mutation probability is as in the \emph{Base} model.

The \emph{Group} model introduces a further process, called ``migration'', working side to side with imitation. Migration regards only $M_g$ and concerns the whole population of agents. At the end of each period, one of the agents with the lowest payoff is selected  as ``migrant''. The ``migrant'' changes its group and adopts the $M_g$ sequence of one of the agents with the highest payoff. Again a mutation probability $\mu$ exists, modeling the individual decision to form a new group. Note that making the mutation probability of migration different from the  imitation one proved not to be  crucial in influencing the final result in some pilot runs. In order to limit the number of parameter in the simulation, we hence decided to maintain a single parameter for both kind of mutations.

\subsection{Results}\label{results}

We explored three different values of the mutation probability parameter and three dimensions of $m_g$, namely $\mu \in \{0.01,0.05,0.10\}$ and $m_g \in \{2,3,4\}$. In order to take into account the stochastic elements of the models, we performed 1000 runs for each parameter configuration. We hence tested 3 parameter configurations for the \emph{Base} model and $3 \times 3$ parameter configurations for the \emph{Group} one, for a total of 12000 runs. Each run lasted for 5000 periods and included 256 agents.

\begin{figure}[p]%
\centering
\includegraphics[width=.325\textwidth]{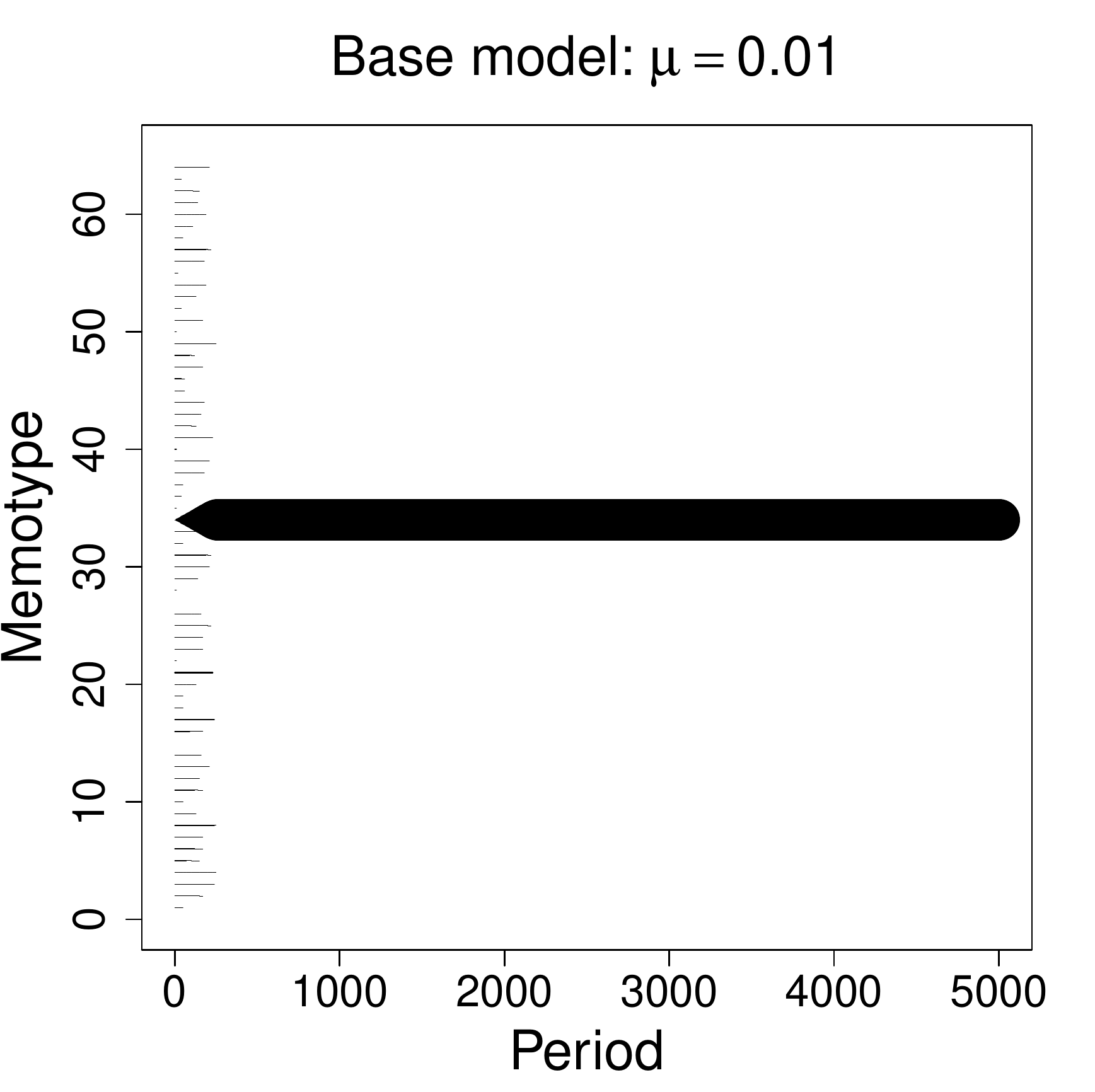} \includegraphics[width=.325\textwidth]{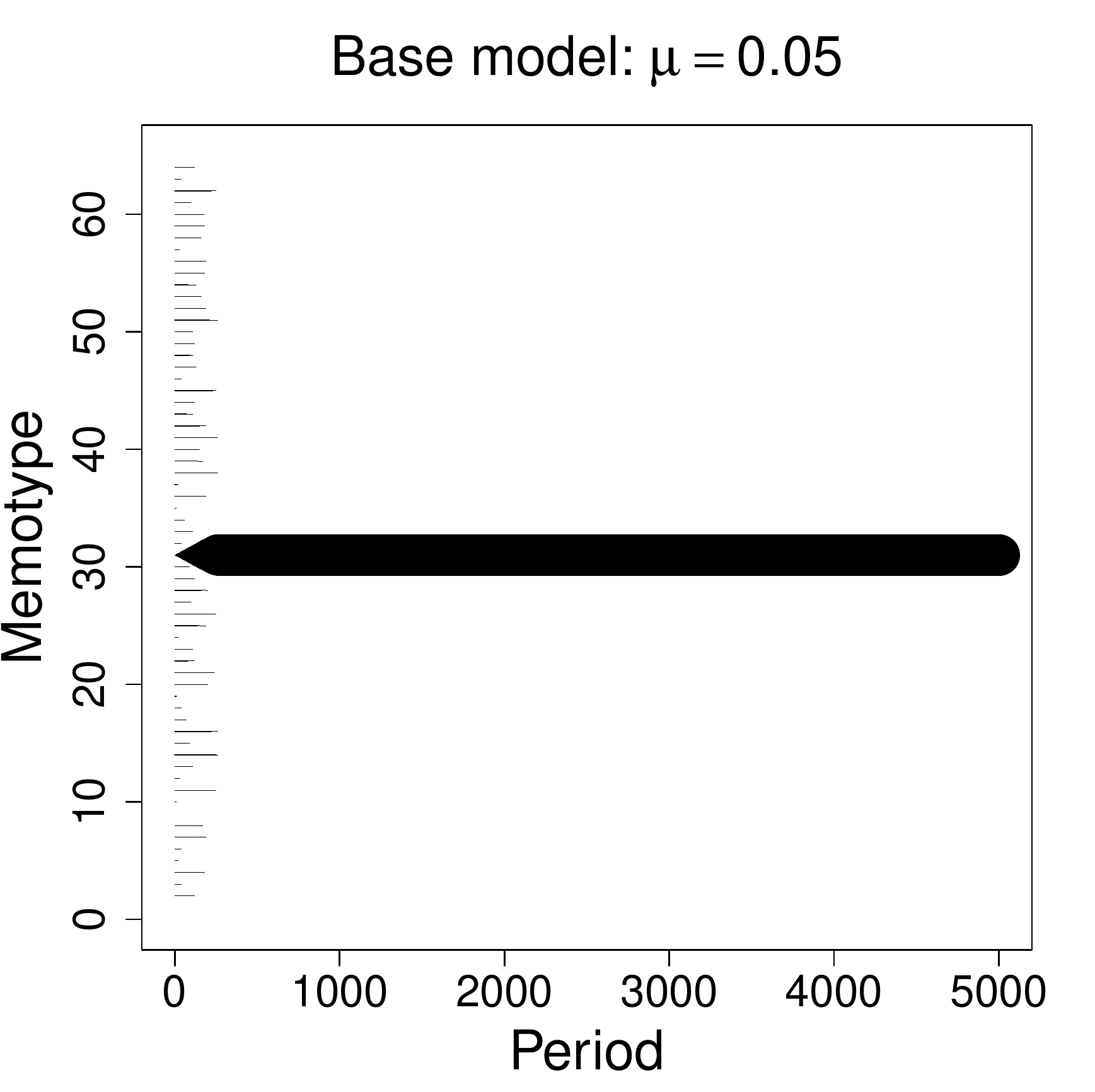} 
\includegraphics[width=.325\textwidth]{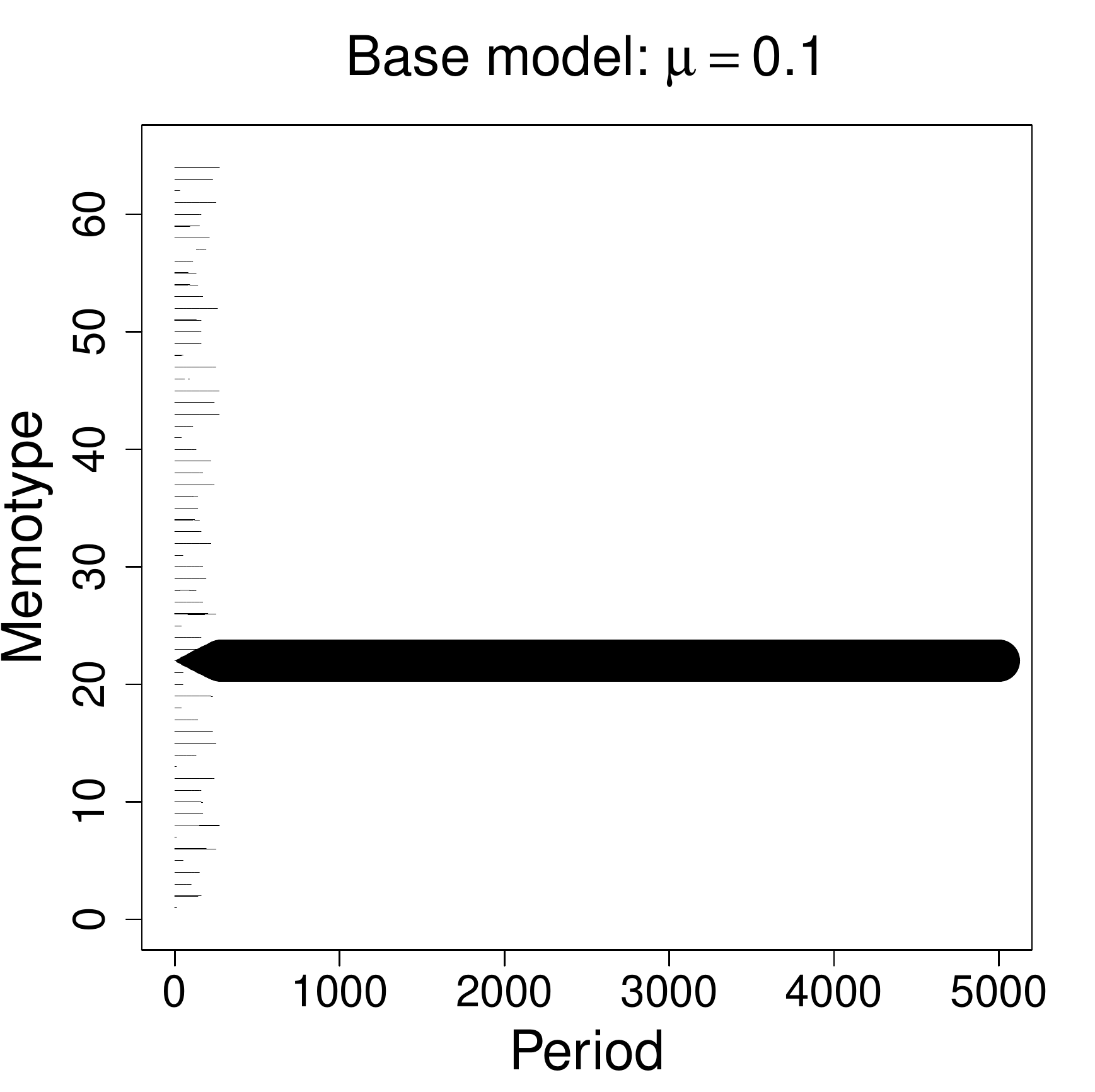}\\
\includegraphics[width=.325\textwidth]{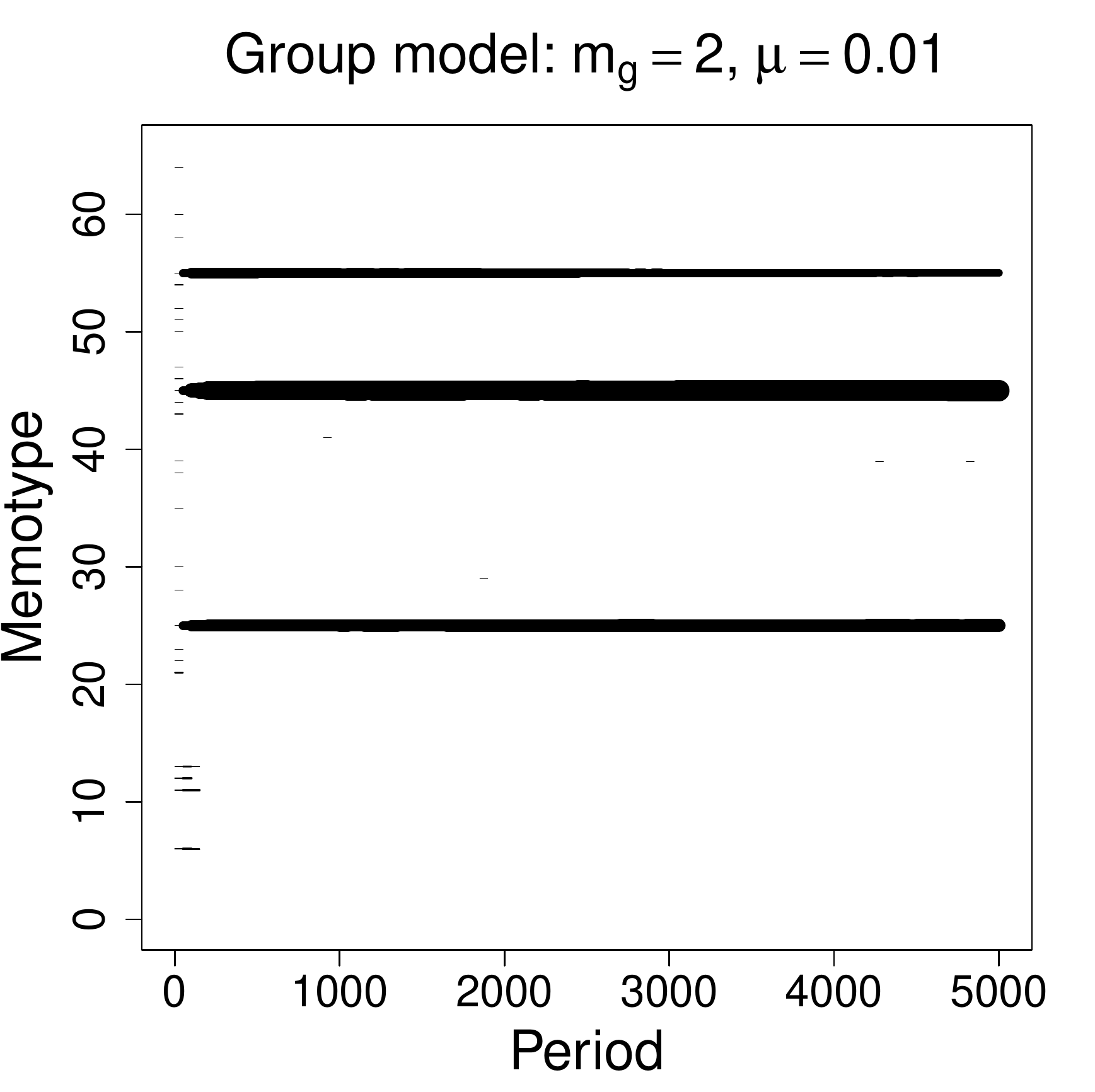} \includegraphics[width=.325\textwidth]{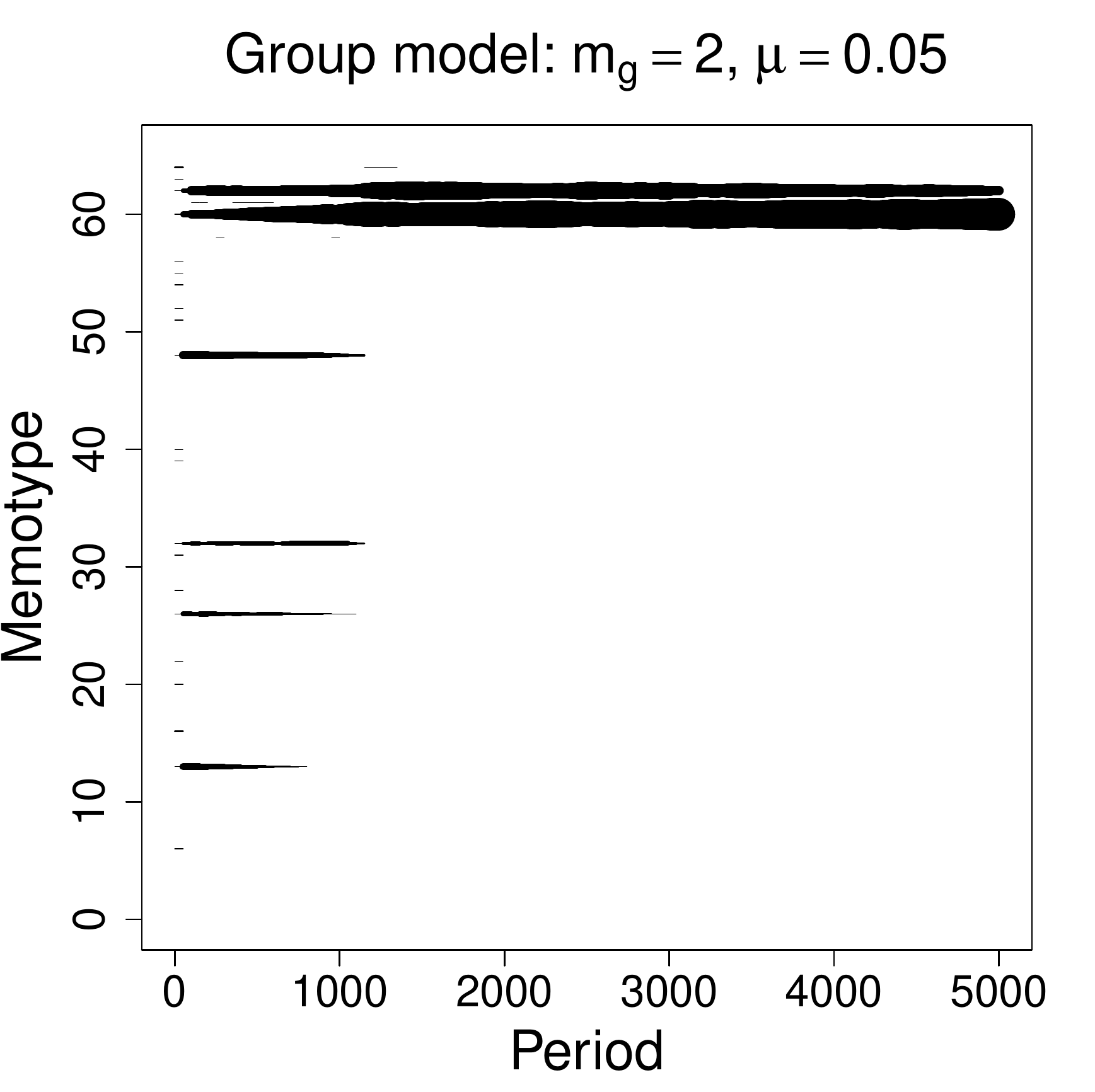} 
\includegraphics[width=.325\textwidth]{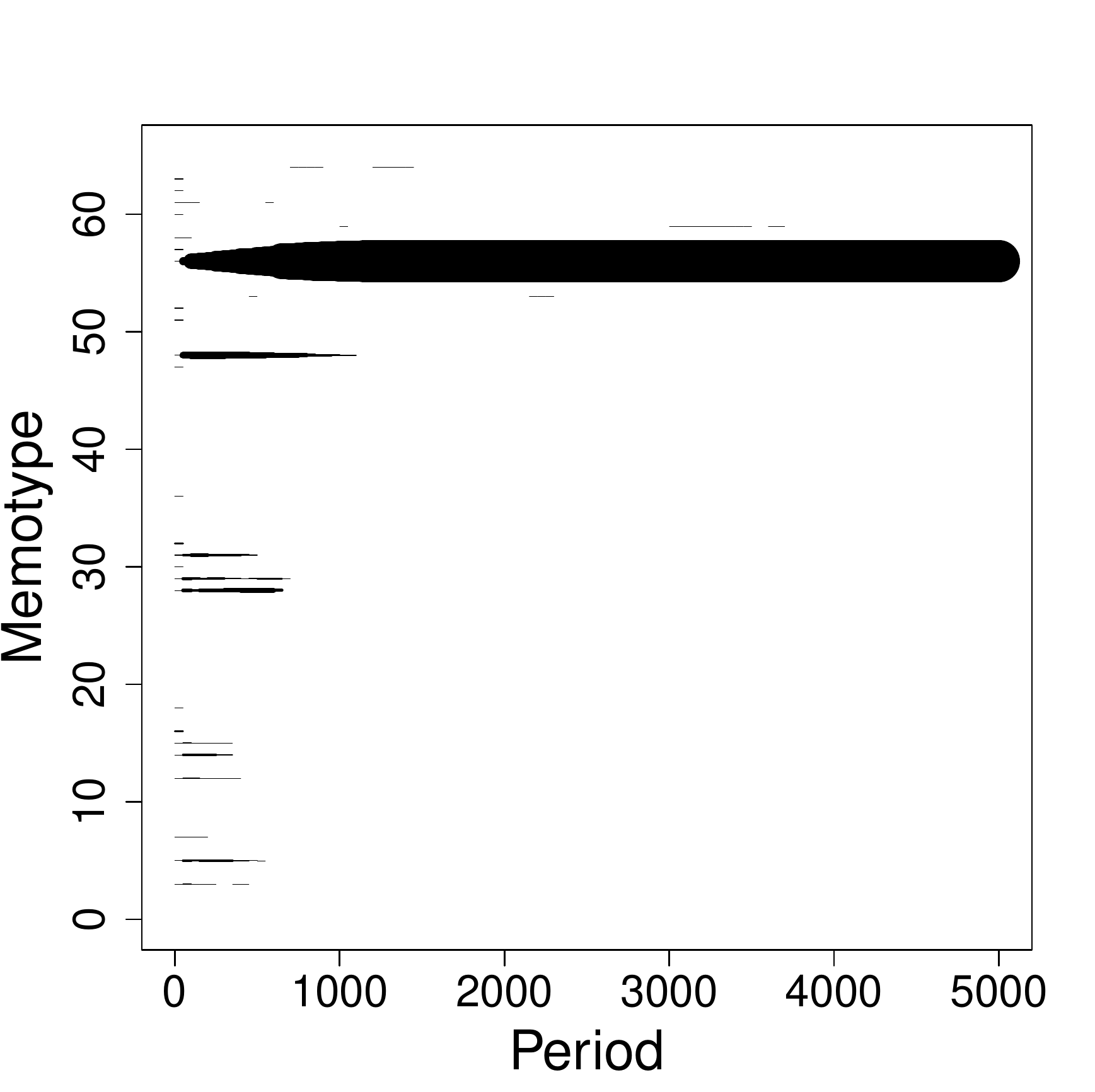}\\
\includegraphics[width=.325\textwidth]{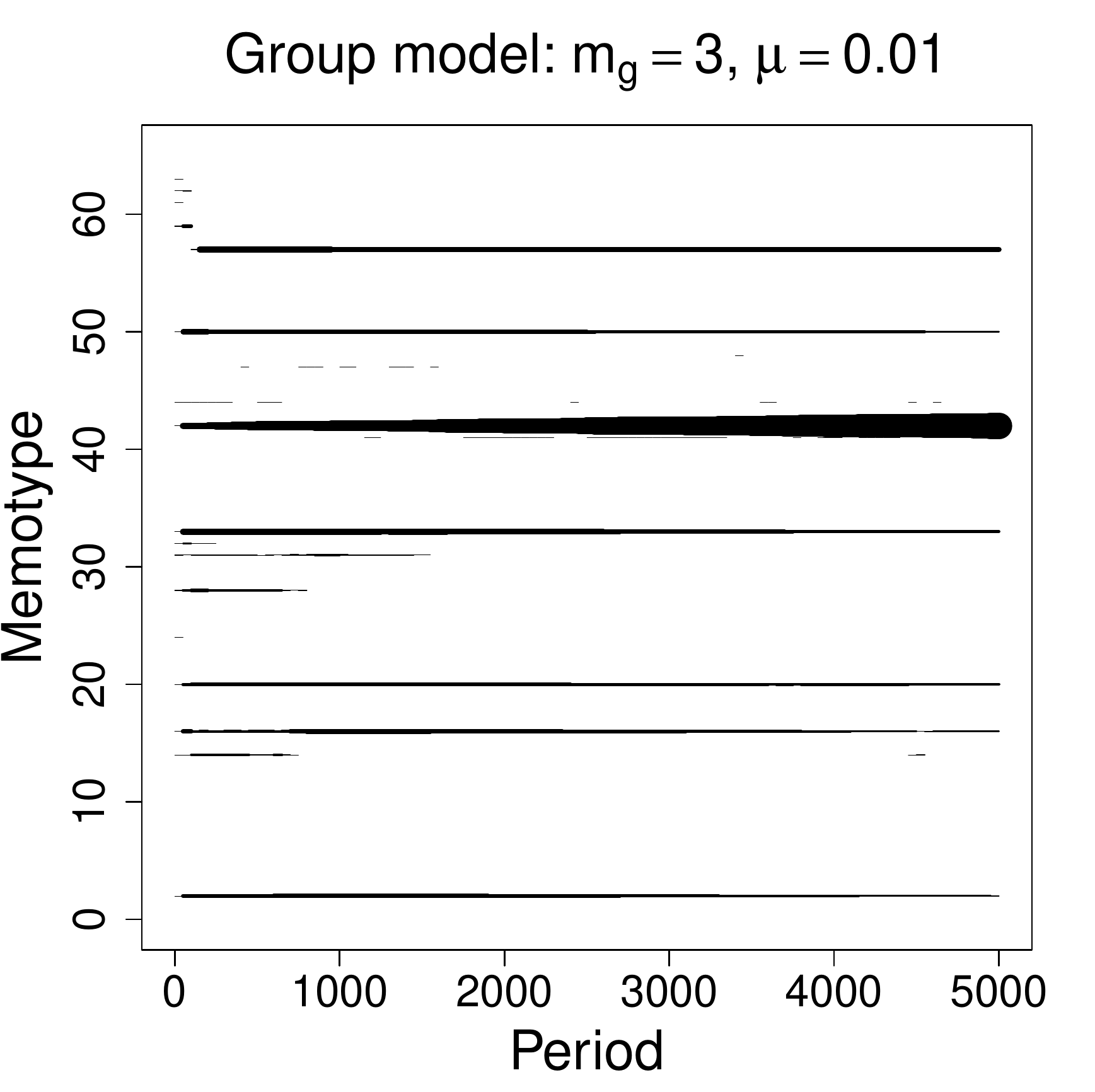} \includegraphics[width=.325\textwidth]{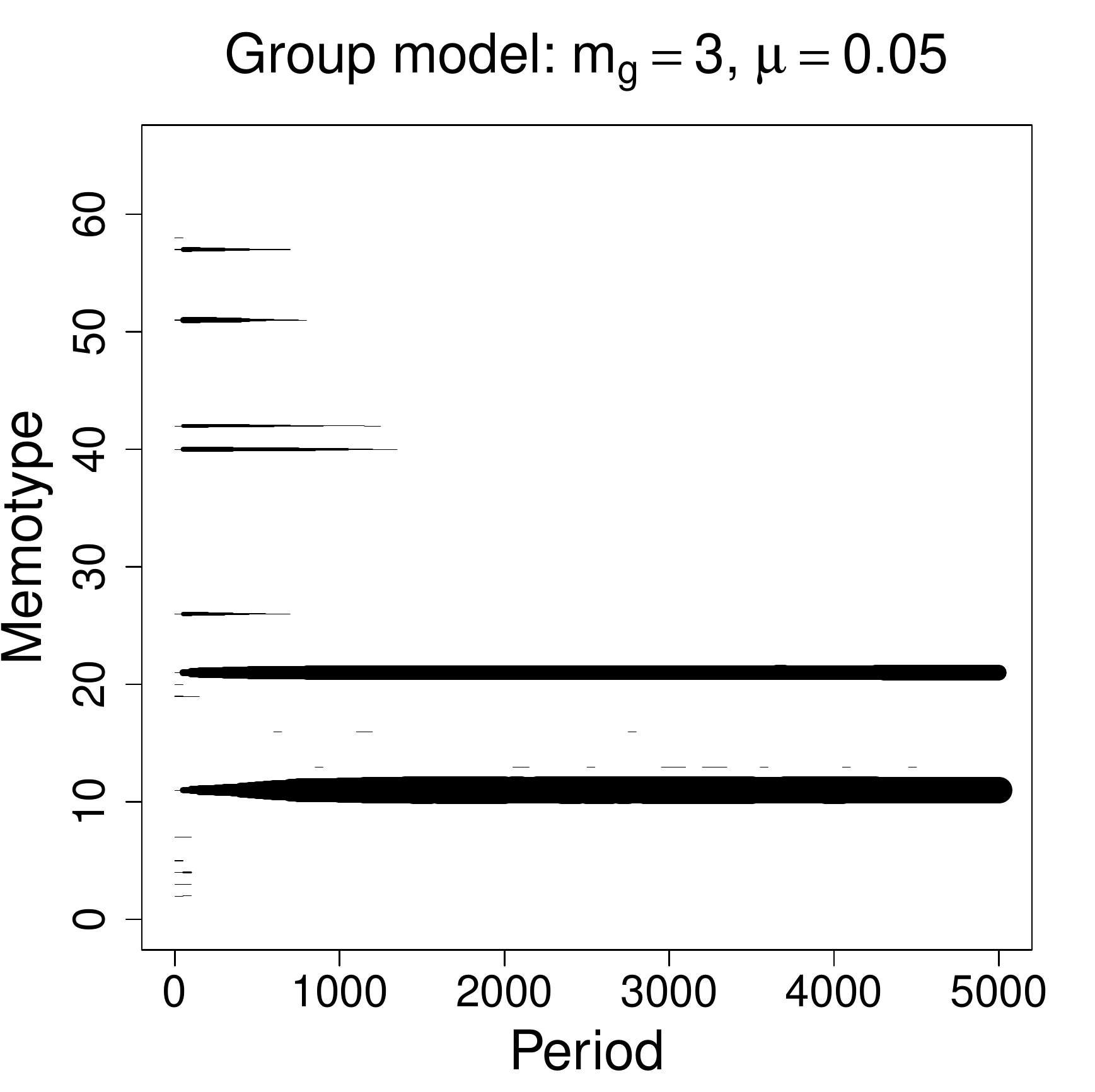} 
\includegraphics[width=.325\textwidth]{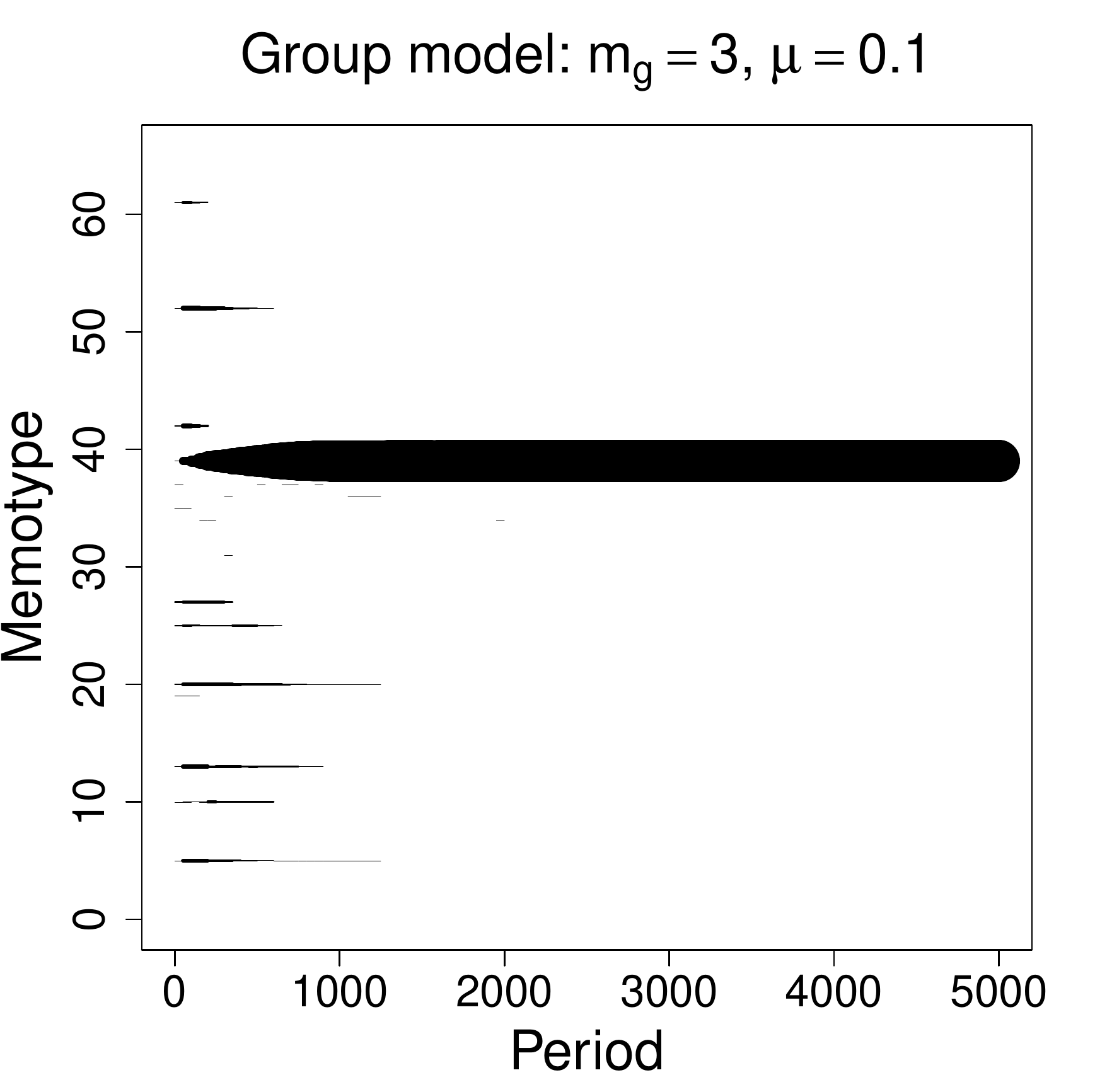}\\
\includegraphics[width=.325\textwidth]{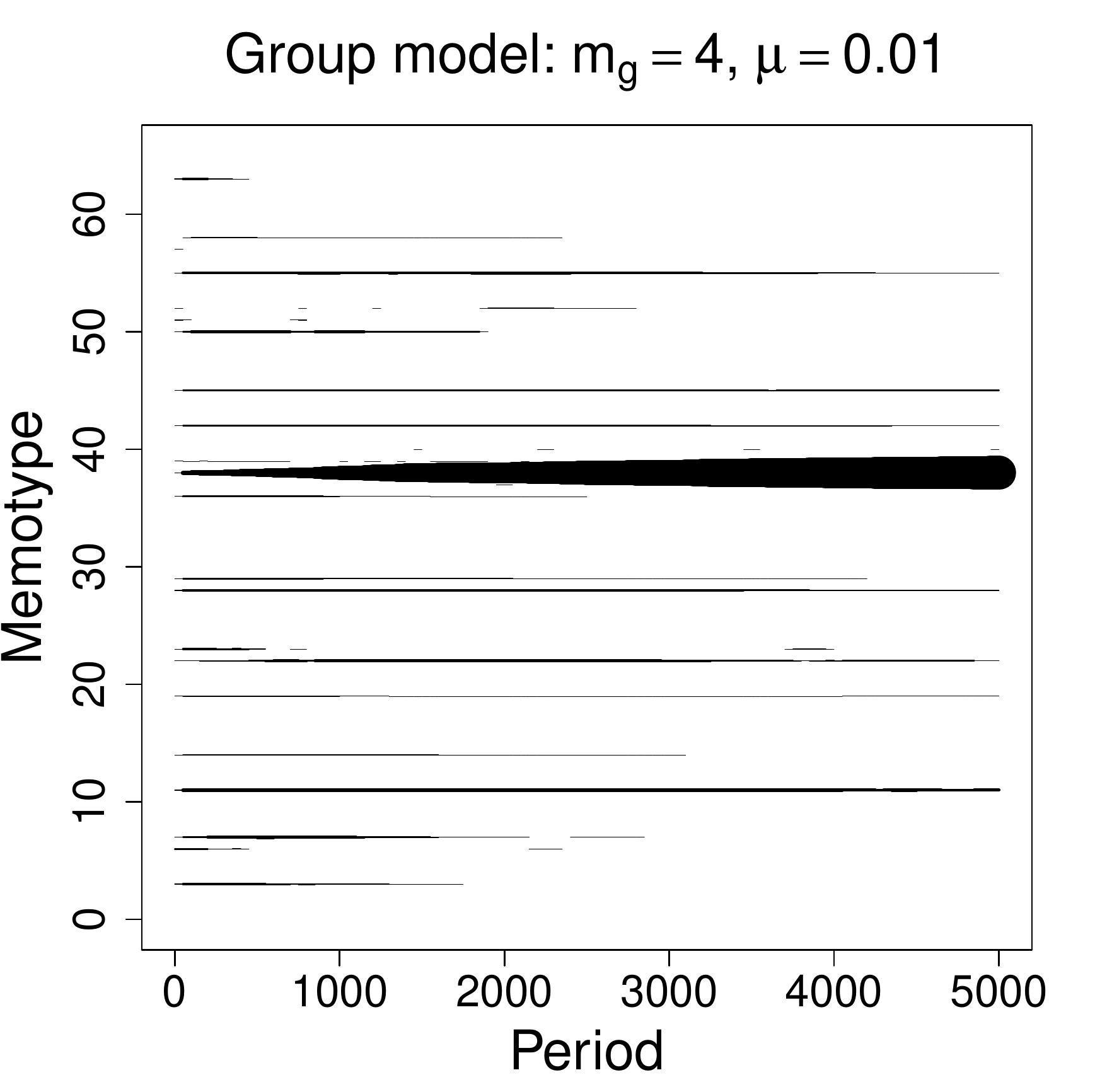} \includegraphics[width=.325\textwidth]{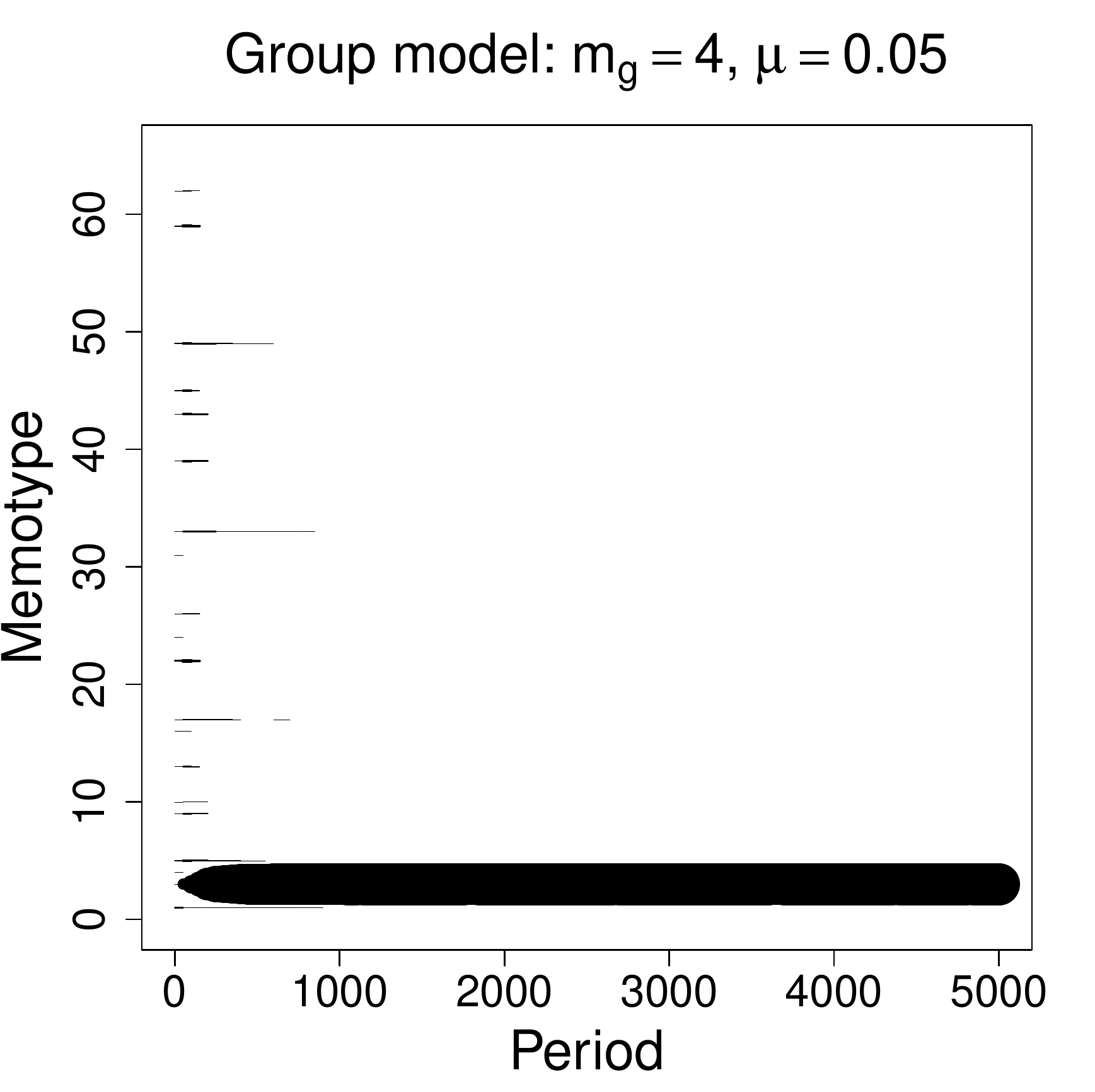} 
\includegraphics[width=.325\textwidth]{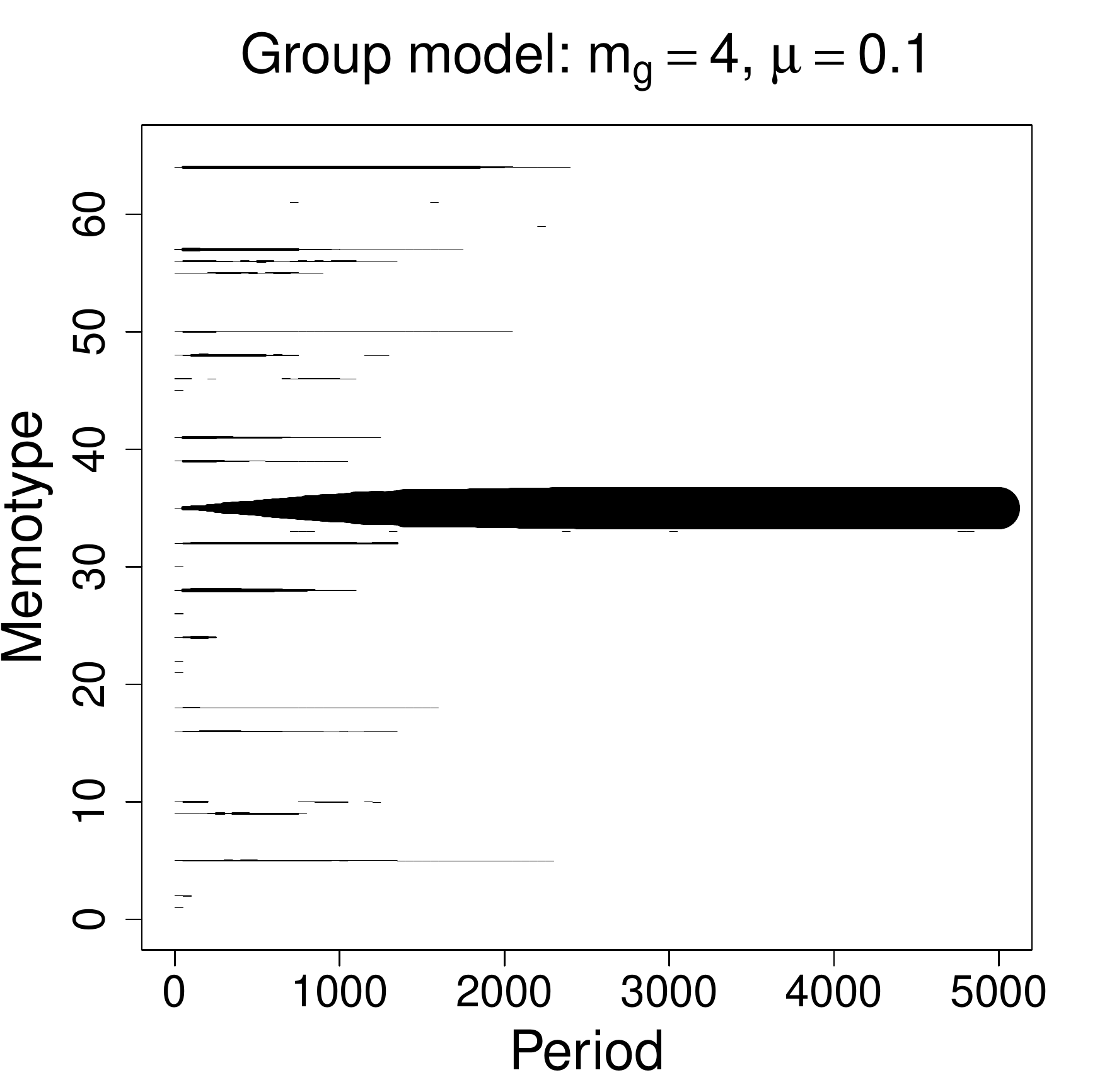}\\
\caption{Agents in each memotype in typical runs of the \emph{Base} and the \emph{Group} models. The line width is proportional to the dimension of the memetic group.}%
\label{fig:typical}%
\end{figure}

Figure \ref{fig:typical} presents the plots resulting from typical runs of the \emph{Base} and \emph{Group} models under different parameter conditions. It is clear than, while the \emph{Base} model tends to produce a single large memetic group (i.e. a group of agents sharing the same memotype), the \emph{Group} one is more favorable to the co-existence of different ``cultures''. However, this holds mainly for low values of $\mu$, while higher ones tend to drive the \emph{Group} model close to the results of the \emph{Base} one.

In order to test whether the differences highlighted in the plots are statistically relevant, we counted the number of memetic groups in each of the 1000 runs performed for each parameter configuration. More specifically, we counted the number of memetic groups composed by more than 4 agent --- i.e a group larger than what we should expect if the agents' memotypes were randomly distributed --- and composed by more than 25 agents, --- i.e. a group including more than 10\% of the agents. A summary of the results is reported in Table \ref{tab:stats140}. Since at the beginning of each run the agents are randomly distributed among memotypes, the reported results refer only to the second half of the simulation, namely from period 2500 onwards.

\begin{table}[t]%
\centering
\begin{tabular}{cccr@{ }lr@{ }lr@{ }l}
\toprule
 & & \multicolumn{7}{c}{Model}\\
\cmidrule(lr){3-9}
 Memetic groups & $\mu$ & \emph{Base} &  \multicolumn{2}{c}{\emph{Group}: $m_g=2$} &  \multicolumn{2}{c}{\emph{Group}: $m_g=3$} &  \multicolumn{2}{c}{\emph{Group}: $m_g=4$} \\
\midrule
 $>4$  & 0.01 & 1.76 (1.20) & 3.20 (0.77) & $^{***}$ &  4.12 (1.65)& $^{***}$ &  3.66 (2.38) & $^{***}$ \\
 $>4$  & 0.05 & 1.84 (1.26) & 2.03 (0.81) & $^{***}$ &  1.80 (0.88) & & 1.77 (0.97) & \\
 $>4$  & 0.10 & 1.88 (1.39) & 1.52 (0.61) & $^{***}$ &  1.39 (0.55) & $^{***}$ &  1.35 (0.57) & $^{***}$ \\
\midrule  
 $>25$ & 0.01  & 1.56 (0.82) & 2.70 (0.75) & $^{***}$ &  2.12 (0.94) & $^{***}$ &  1.67 (0.87) & $^{**}$ \\
 $>25$ & 0.05  & 1.61 (0.84) & 1.68 (0.67) & $^{*}$ &  1.44 (0.62) & $^{***}$ &  1.40 (0.58) & $^{***}$ \\
 $>25$ & 0.10  & 1.62 (0.90) & 1.36 (0.51) & $^{***}$ &  1.28 (0.46) & $^{***}$ &  1.22 (0.41) & $^{***}$ \\
\bottomrule
\end{tabular}
\caption{Number of memetic groups including more than 4 and 25 agents in the final 2500 periods of the simulation. Results are averaged over 1000 runs per experimental condition with standard deviations in parenthesis. Significance codes ($t$ test between corresponding conditions of the \emph{Base} and the \emph{Group} models):  $^{***}$ $p<0.001$, $^{**}$ $p<0.01$,  $^{*}$ $p<0.05$.}
\label{tab:stats140}
\end{table}

The quantitative analysis confirms the hints offered by Figure \ref{fig:typical}. The \emph{Group} model leads to a significantly higher number of coexisting memetic groups than the \emph{Base} one, at least for $\mu=0.01$.  A $t$ test shows indeed that, for all $m_g$ values, the number of memetic groups is significantly higher in the \emph{Group} model than in the corresponding \emph{Base} model. The standard for the \emph{Group} model is indeed the coexistence of two large groups (more than 25 agents) plus possibly one or two smaller groups. On the other hand, only one large group is almost ever present int the \emph{Base} model, leaving at most the space for one further smaller group.

The picture changes when $\mu$ increases. From one hand, the number of memetic groups in the \emph{Base} model tends to increase: a fact that is rather surprising, since high mutation rates weaken the ``efficiency'' of the selection process, leaving more room for alternative memotypes. On the other hand, increasing the mutation probability has rather dramatic effects for the \emph{Group} model. For any value of $\mu \geq 0.05$ the number of memetic groups  in the \emph{Group} model is no longer higher than the one in the \emph{Base model}. For $\mu = 0.10$ this figure becomes even smaller than the corresponding one in the \emph{Base} model.

\section{The niche construction models}\label{coevolution}

The models presented above offer some interesting insights, but only partially answer to our questions. This for two main reasons. First, while they succeed in modeling the selection process, they do not reproduce the death and birth of groups typical of the cultural world. In other words, they do not lead to a real evolution of alternative cultures.  Second, the \emph{Group} model lead to the coexistence of different memotypes only under the condition of high copy fidelity of cultural traits: something that is likely not to happen in the real world \citep{RB2005}.

What is probably missing from these models is that organisms often modify the environment where they live, affecting also the possibilities of making a living of other organisms. Some authors further expand this idea by arguing that, since the action of any organism modifies the availability of resources for other organisms, this changes the selective pressures existing in the environment. This process is called ``niche construction'' \citep{Laland2008,Odling-Smee2003} and there is no reason to imagine that it does not apply to cultural evolution. On the contrary, modern human economies represent a clear example of enormous differentiation based on what other individuals do, think, produce and consume. In order to reproduce this and similar facts, we hence decided to build two further models where the environment is no longer fixed, but varies throughout the simulation as a function of the agents' memotypes.

\subsection{Model definiton}

The new models derive from the previous ones and maintain their structure. More specifically, everything but the environment definition remains as in the \emph{Base} and the \emph{Group} models. We hence called the two new models \emph{NC-base} and \emph{NC-group}, where ``NC'' stands for niche construction. What changes is that the environment definition is no longer a process on its own. The environment is instead built by concatenating the complements of each element of the agents' memotypes. An example will make this point clearer. Given a model including three agents with memotypes $M_1 =\{0,0,0\}$, $M_2=\{1,1,1\}$ and $M_3=\{1,0,1\}$, the environment is $E=\{1,1,1,0,0,0,0,1,0\}$, which leads to a payoff of two, one and zero points respectively for the three agents. The choice of using the memotype complements permits to avoid the tautological condition where agents are adapted to themselves. On the contrary, each agent creates the possibility of making a living (a niche) for a different agent in a dynamic process that mimics what happens in the natural and the social world.

\subsection{Results}

\begin{figure}[p]%
\centering
\includegraphics[width=.325\textwidth]{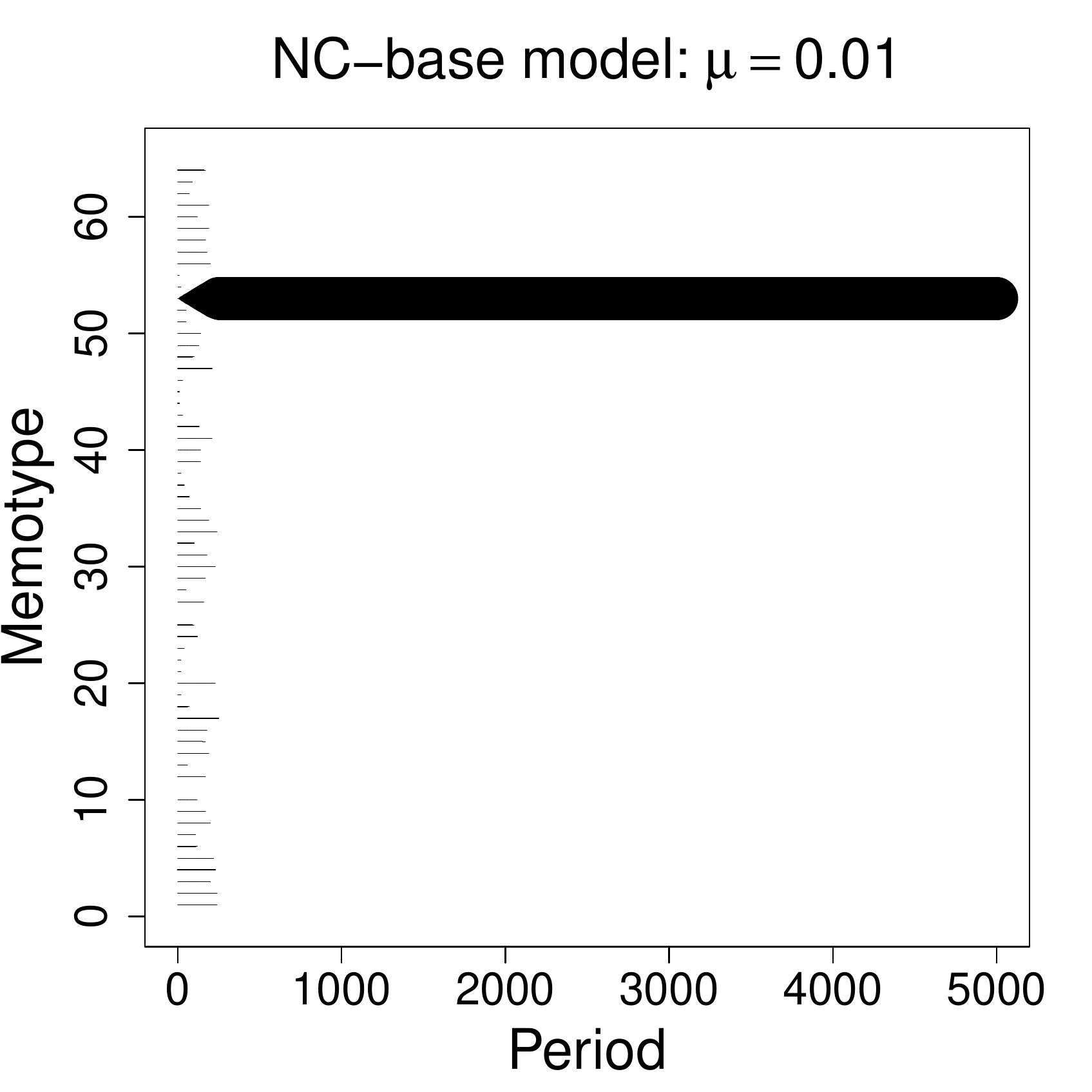} \includegraphics[width=.325\textwidth]{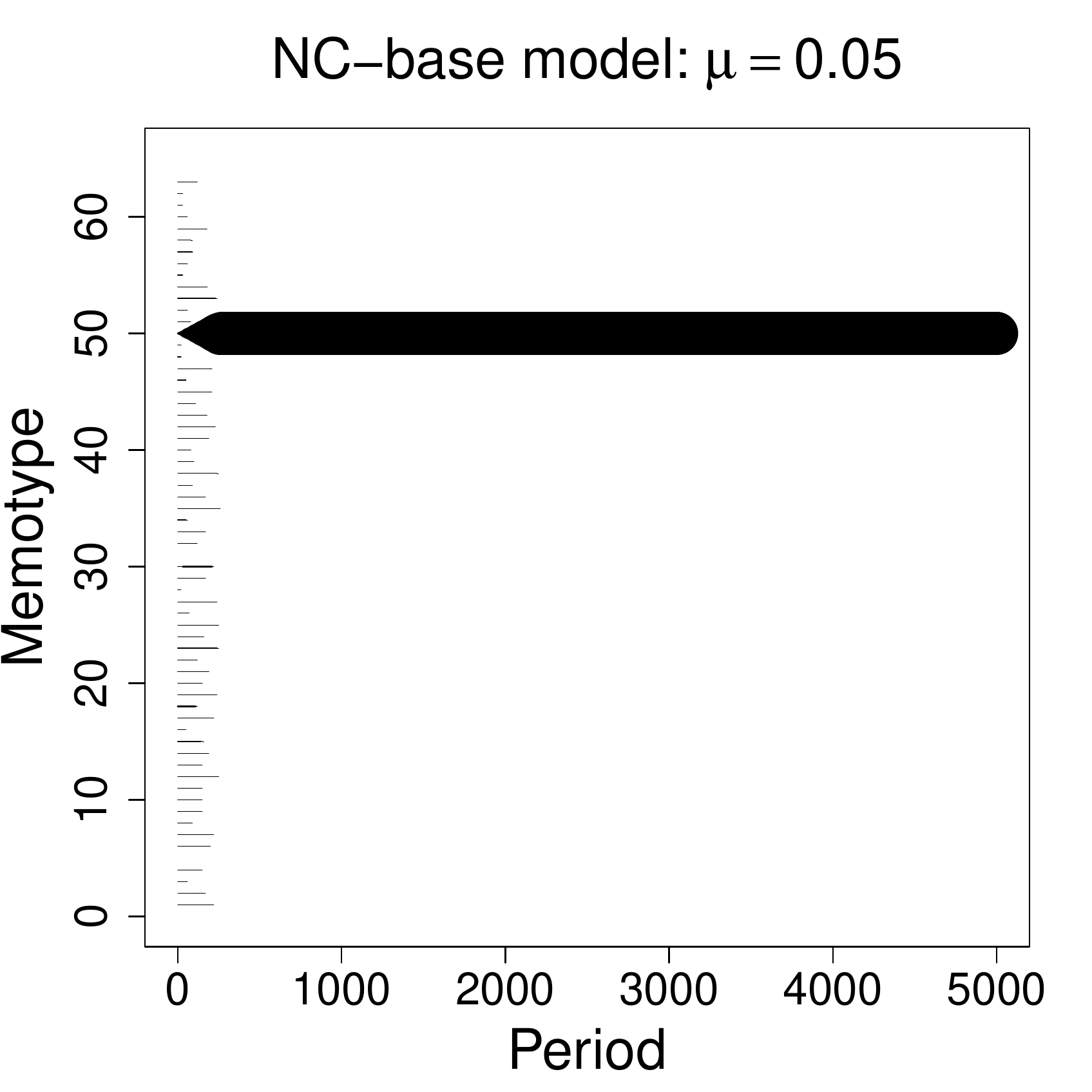} 
\includegraphics[width=.325\textwidth]{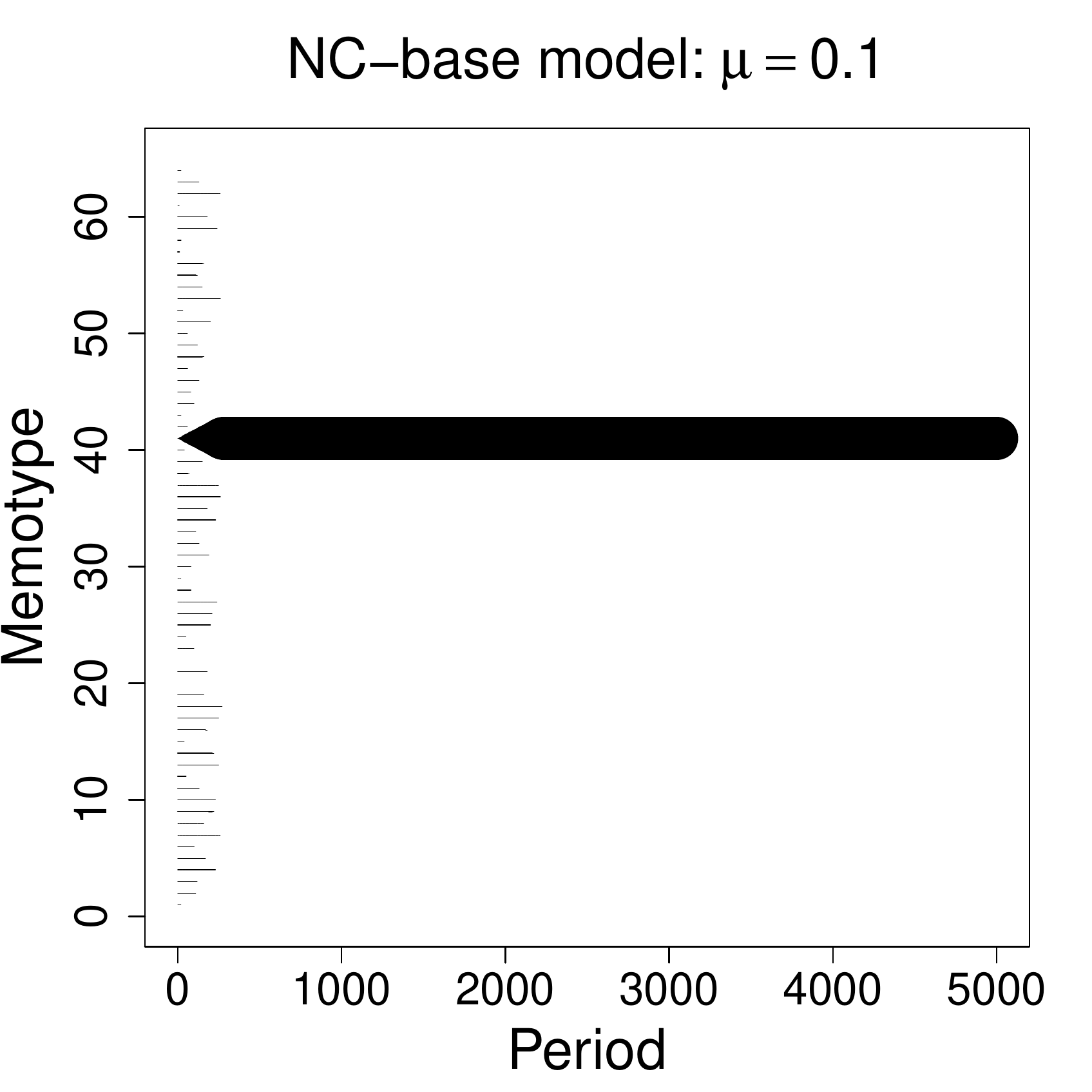}\\
\includegraphics[width=.325\textwidth]{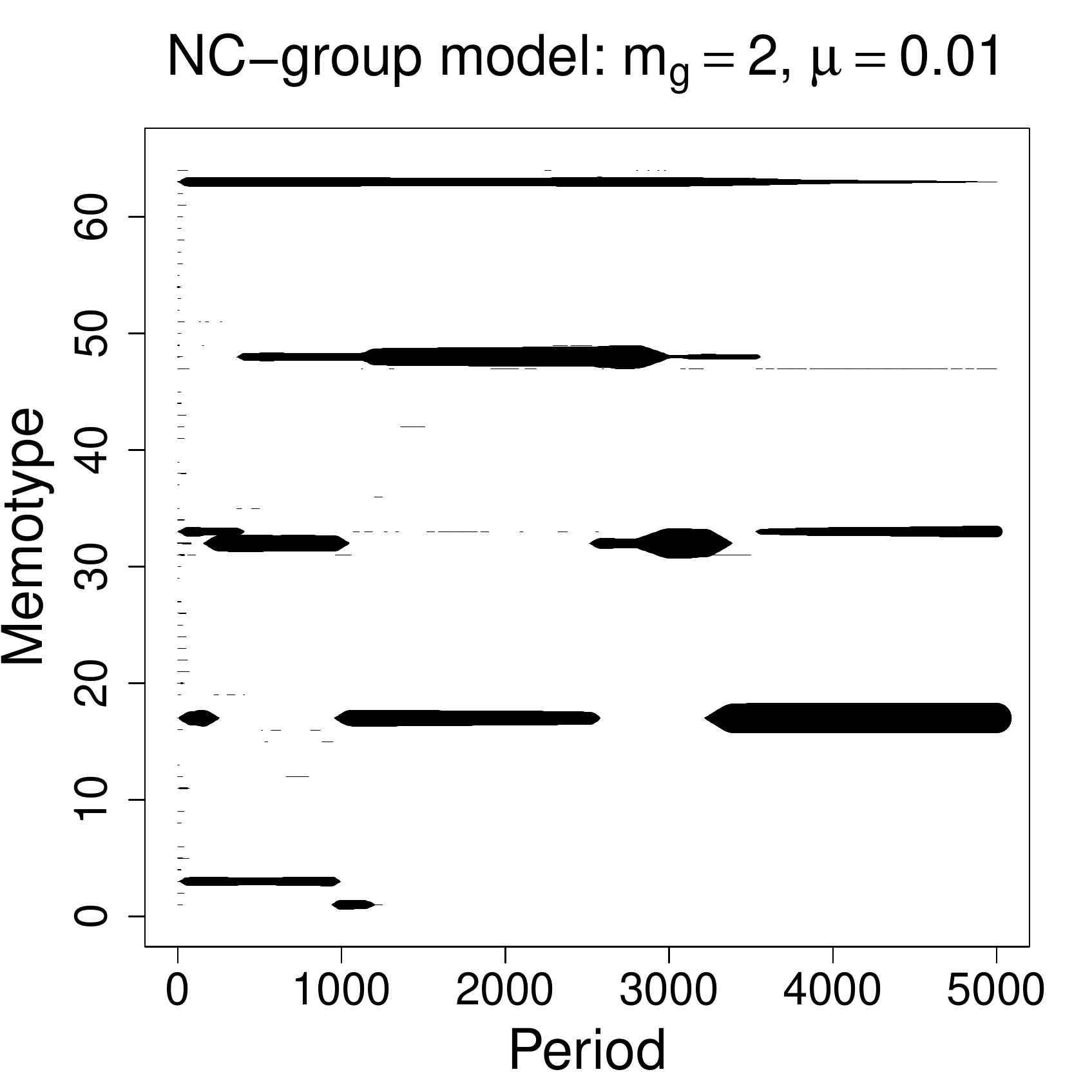} \includegraphics[width=.325\textwidth]{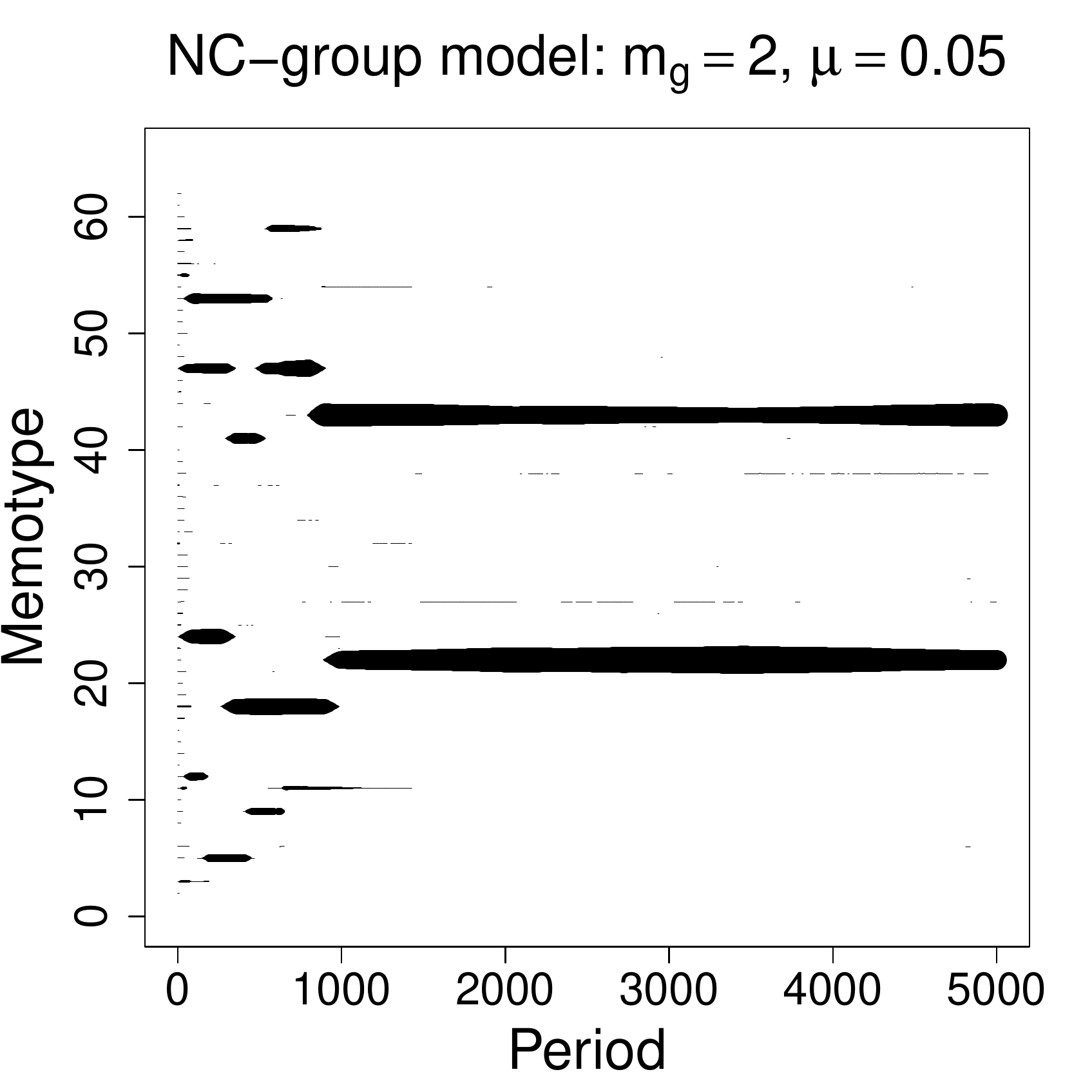} 
\includegraphics[width=.325\textwidth]{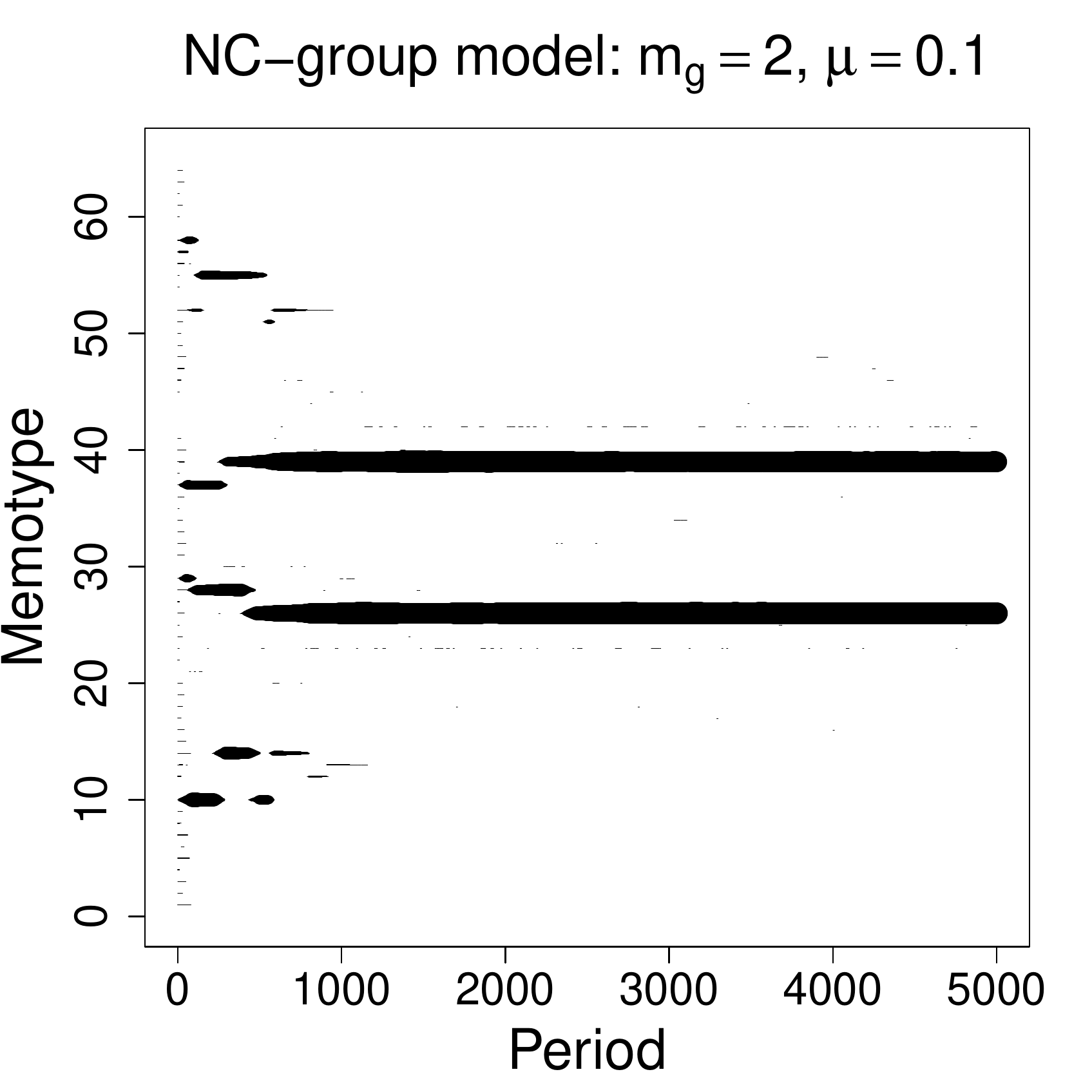}\\
\includegraphics[width=.325\textwidth]{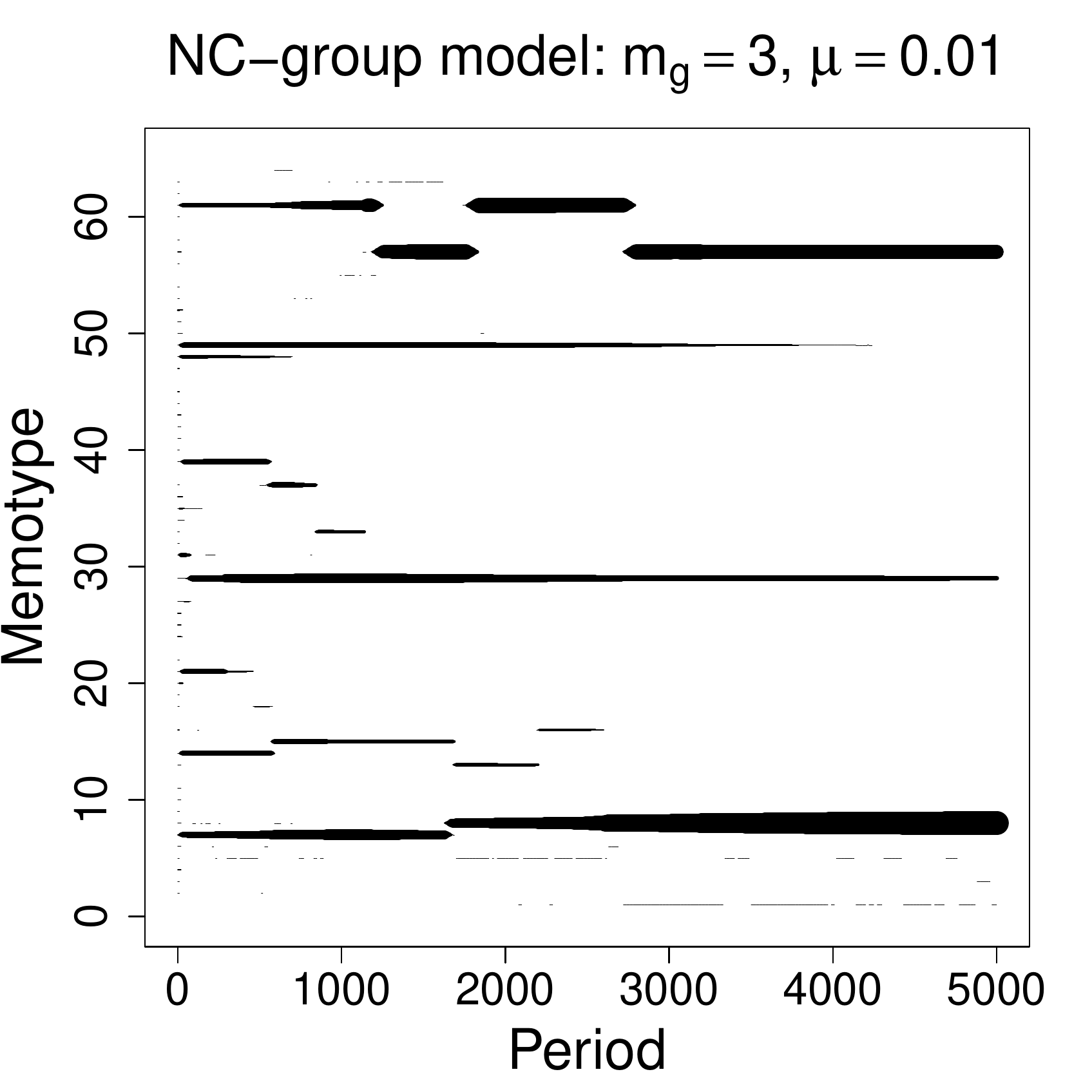} \includegraphics[width=.325\textwidth]{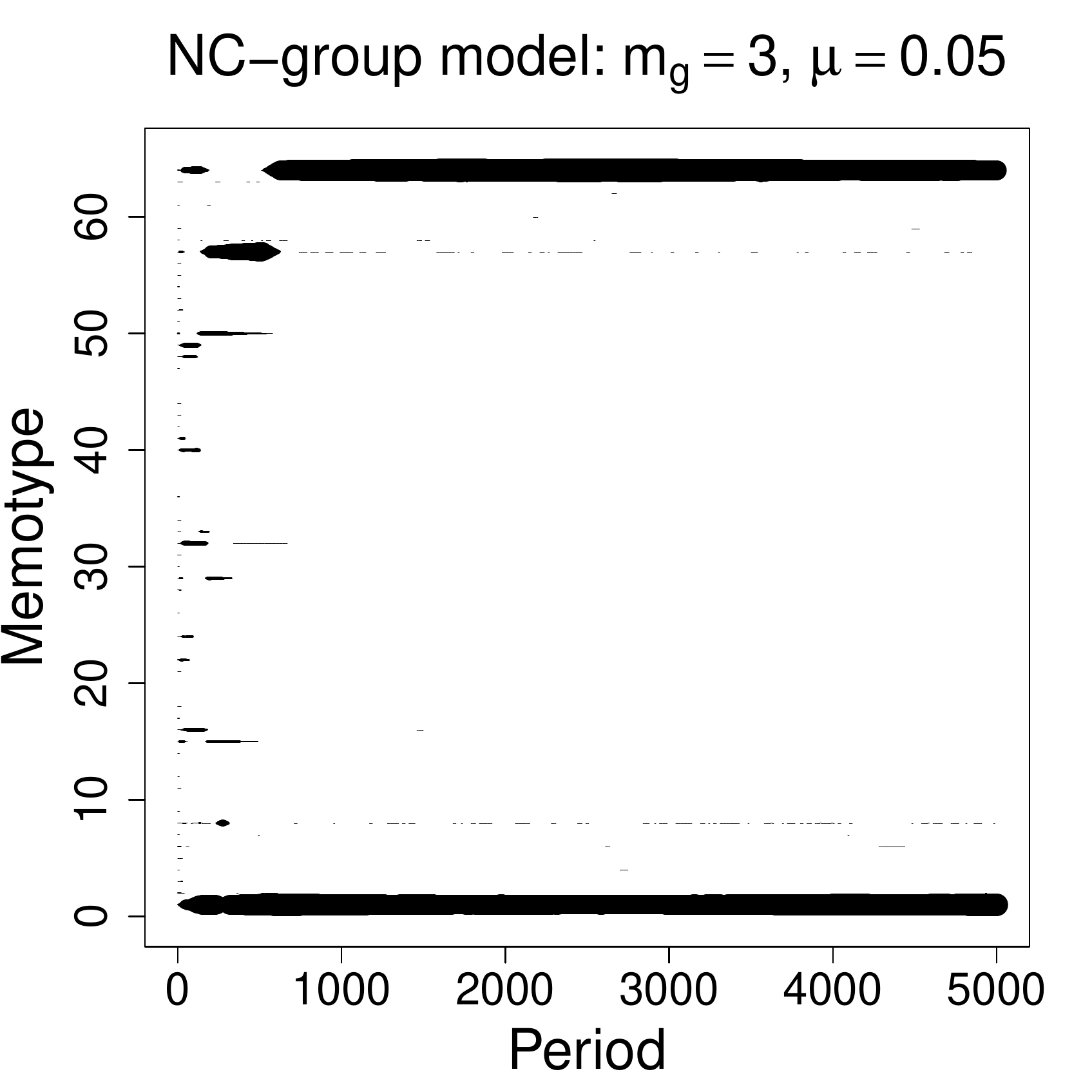} 
\includegraphics[width=.325\textwidth]{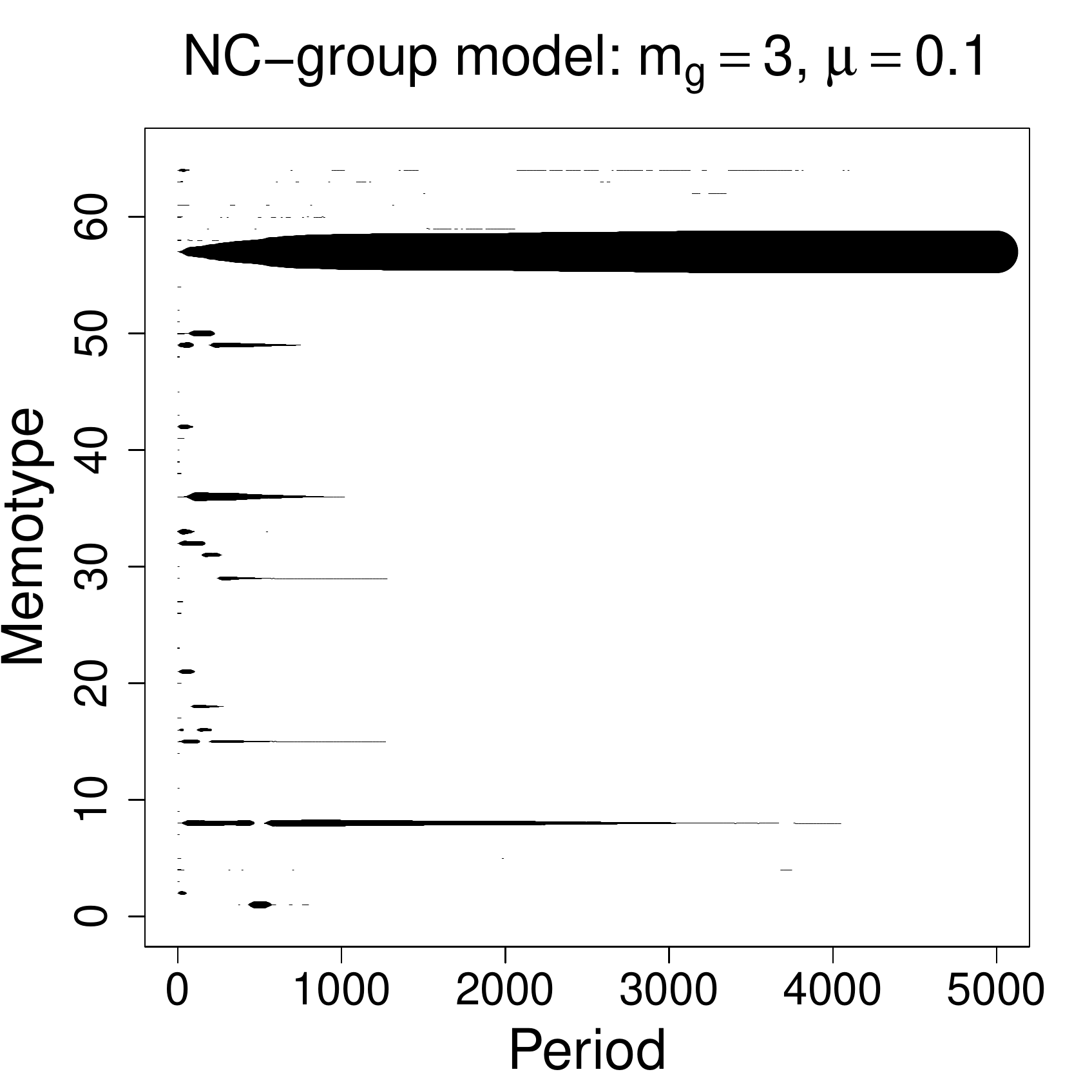}\\
\includegraphics[width=.325\textwidth]{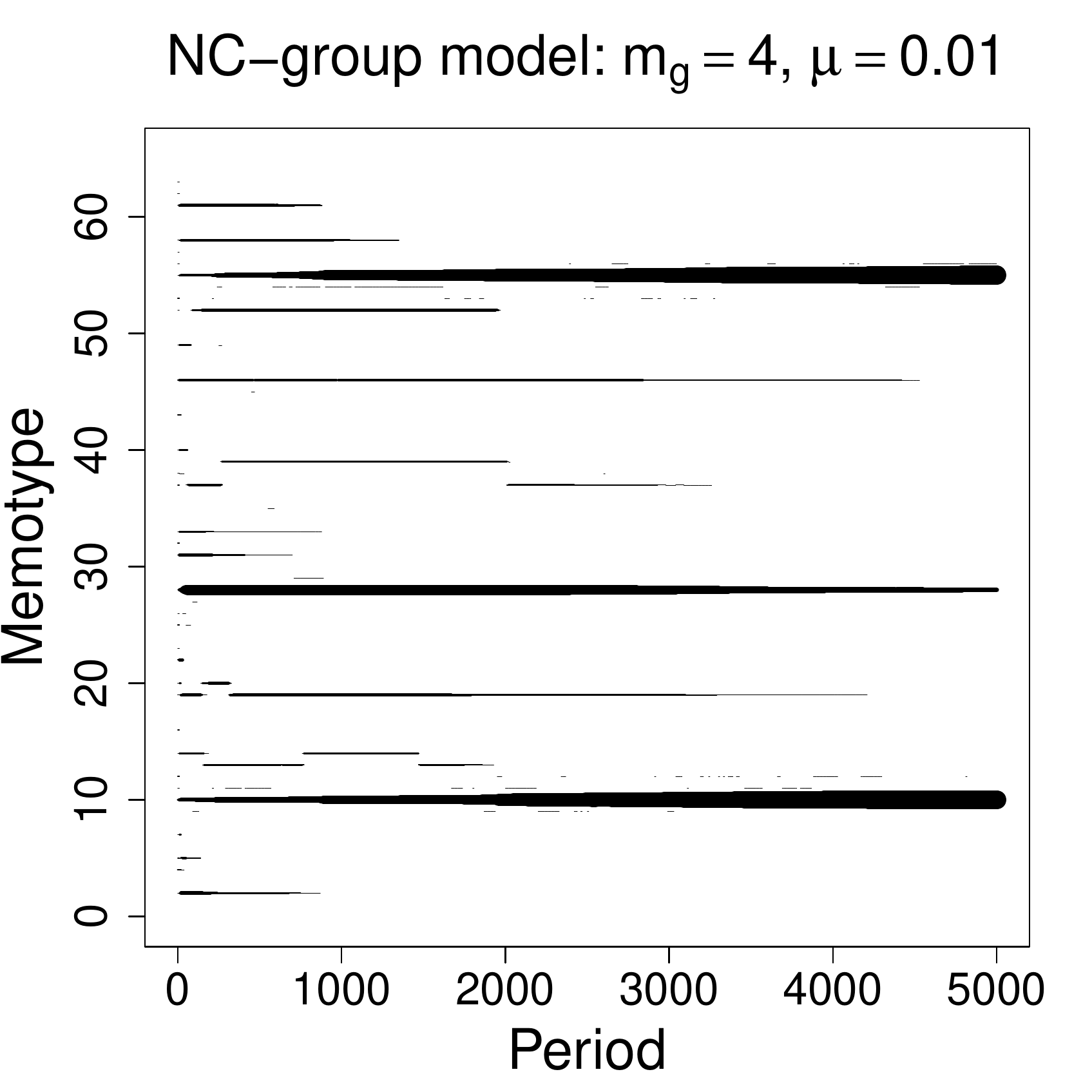} \includegraphics[width=.325\textwidth]{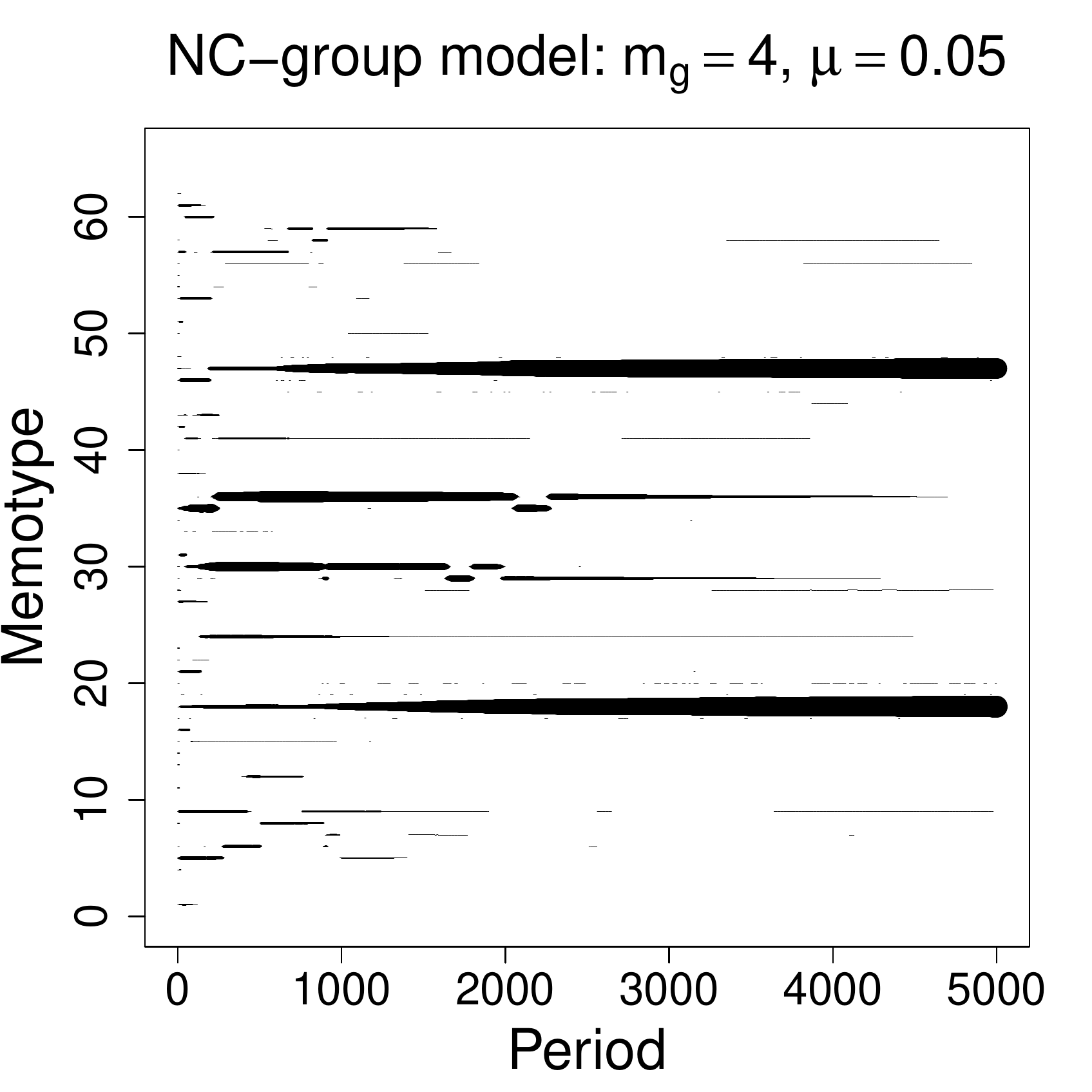} 
\includegraphics[width=.325\textwidth]{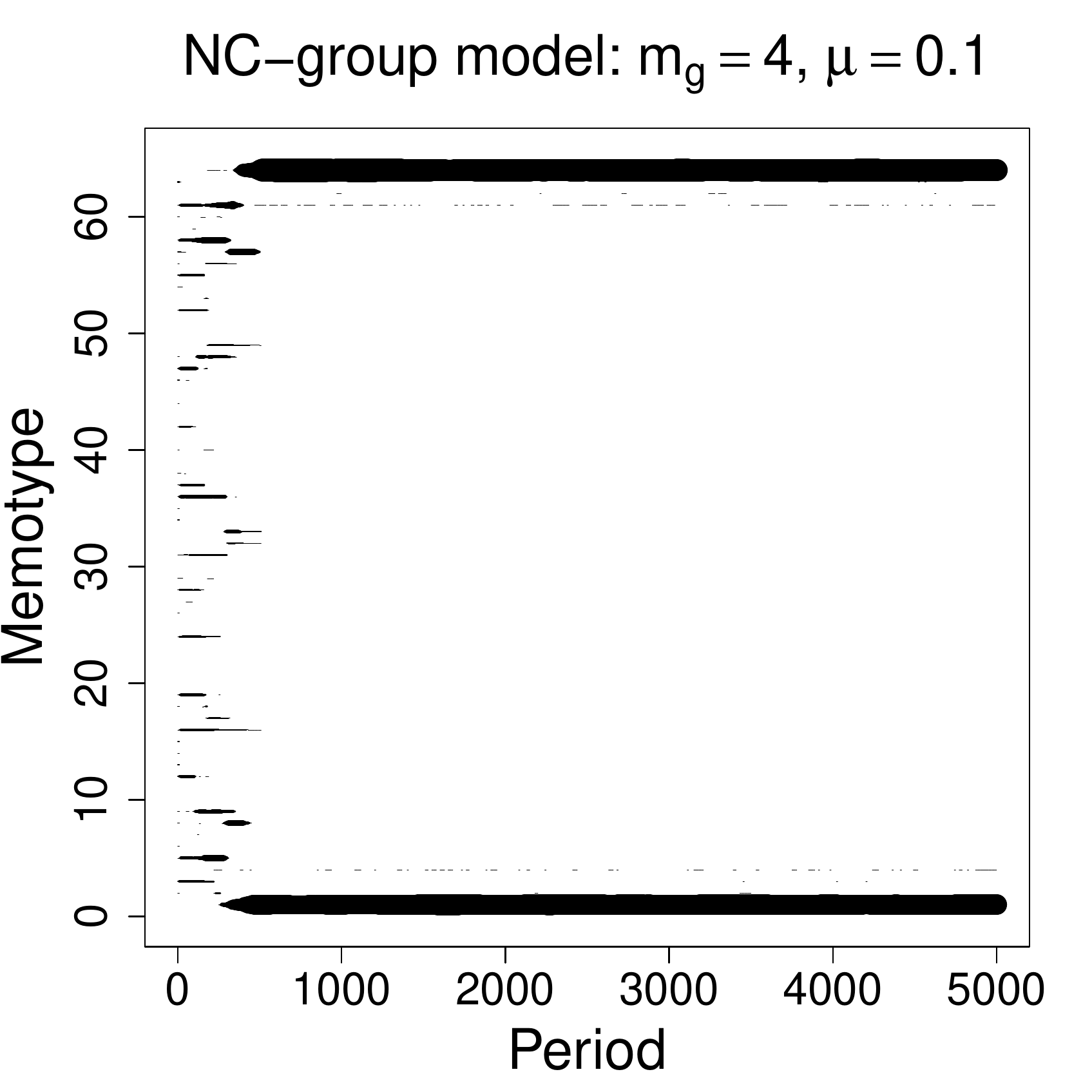}\\
\caption{Agents in each memotype in typical runs of the \emph{NC-base} and the \emph{NC-group} models. The line width is proportional  the dimension of the memetic group.}%
\label{fig:niches}%
\end{figure}

In order to test the new models we used the same parameter configurations explored for the cultural evolution models (except, obviously, for the environment length which is now given by the number of agents times the memotype length). We hence performed a total of 12000 runs of the new models. Figure \ref{fig:niches} presents the plots of typical runs of the \emph{NC-base} and \emph{NC-group} models under different parameter conditions while Table \ref{tab:statsNiches} summarizes the statistics.

\begin{table}[t]%
\centering
\begin{tabular}{cccr@{ }lr@{ }lr@{ }l}
\toprule
 & & \multicolumn{7}{c}{Model}\\
\cmidrule(lr){3-9}
Memetic groups & $\mu$ & \emph{NC-base} &  \multicolumn{2}{c}{\emph{NC-gr.}: $m_g=2$} &  \multicolumn{2}{c}{\emph{NC-gr.}: $m_g=3$} &  \multicolumn{2}{c}{\emph{NC-gr.}: $m_g=4$} \\
\midrule
 $>4$  & 0.01 & 1.31 (0.66) & 2.46 (0.73) & $^{***}$ &  2.57 (1.03) & $^{***}$ & 2.70 (1.11) & $^{***}$ \\
 $>4$  & 0.05 & 1.30 (0.62) & 2.02 (0.25) & $^{***}$ &  2.00 (0.31) & $^{***}$ & 2.16 (0.41) & $^{***}$ \\
 $>4$  & 0.10 & 1.30 (0.65) & 2.12 (0.08) & $^{***}$ &  1.92 (0.25) & $^{***}$ & 2.02 (0.15) & $^{***}$ \\
\midrule  
 $>25$ & 0.01  & 1.25 (0.55) & 2.25 (0.57) & $^{***}$ & 2.13 (0.61) & $^{***}$ & 2.14 (0.45) & $^{***}$ \\
 $>25$ & 0.05  & 1.23 (0.49) & 1.99 (0.19) & $^{***}$ & 1.92 (0.31) & $^{***}$ & 2.03 (0.21) & $^{***}$ \\
 $>25$ & 0.10  & 1.23 (0.53) & 2.00 (0.02) & $^{***}$ & 1.88 (0.29) & $^{***}$ & 2.00 (0.13) & $^{***}$ \\
\bottomrule
\end{tabular}
\caption{Number of memetic groups including more than 4 and 25 agents in the final 2500 periods of the simulation. Results are averaged over 1000 runs per experimental condition with standard deviations in parenthesis. Significance codes ($t$ test between corresponding conditions of the \emph{NC-base} and the \emph{NC-group} models):  $^{***}$ $p<0.001$, $^{**}$ $p<0.01$,  $^{*}$ $p<0.05$.}
\label{tab:statsNiches}
\end{table}

From Figure \ref{fig:niches} two main consideration arise. First, the \emph{NC-group} model leads to the coexistence of memetic groups under all the examined parameter conditions  and it is much less weakened by high mutation rates than the \emph{Group} model. Second, the evolutionary process looks more credible, with births and deaths of ``cultures'' and, more generally, with a greater number of changes during the course of the run.

Table \ref{tab:statsNiches} confirms that the number of coexisting memotype groups is significantly larger in the \emph{NC-group} model than in the \emph{NC-base} for all parameter configurations. Note also that standard deviations are much lower than in the cultural evolution models, witnessing a greater convergence of the niche construction models towards a clear equilibrium.

\section{Discussion}\label{discussion}

The main result of research is that group definition is effective in producing cultural heterogeneity, even in condition of a partial permeability of the boundaries. This support our first hypothesis, i.e. that the coexistence among cultures needs (imperfect) boundaries, under less restrictive condition than the Axelrod's model. In the \emph{Base} model coexistence cannot last, since selective pressures drive the whole population towards a single memotype even when alternative niches exists in the environment. The \emph{Group} model limits this process by reducing the flux between different memetic groups. This moves part of the selective pressure within the boundaries of each group, even if imitation and migration tend still  to favor the best memotypes. The creation of group boundaries reduces within-group  variation and increases between-group variation, leading hence to the maintaining of coexisting cultures. It is worth noting that this result supports the cultures-as-species idea advanced by \citet{PM2004}. In order to have cultural differences the memetic flux among groups must be limited, in analogy with the interbreeding limits existing among sexual reproducing organisms that stop the genetic flux and permits speciation.

A second point is that the group section of the memotype ($M_g$) has the value of a common resource for each group. Given the model definition, $M_g$  does not represent an arbitrary tag \citep[as, for instance, in][]{Axelrod1997a}, but possesses an actual adaptive value.\footnote{It is worth noting that some explorations of the model using purely symbolic group definition traits --- i.e. traits that were not checked agaisnt the environment --- did not produce a significant coexistence of cultures. While this is not crucial for the current discussion, it may represent the subject of further research.} Being a part of the memotype checked against the environment, $M_g$ contributes indeed actively to the agents' earnings. Successful group are hence groups with a group memotype well adapted to the current environment. This is equivalent to holding a specific knowledge or capacity in human societies. For instance, Renaissance artisans' guilds often kept secret their techniques in order to preserve the competitive advantage of their members. These techniques both helped to define the guilds themselves and represented the foundations of their wealth. Similarly, in our model $M_g$ both identifies the group and allows a successful exploitation of the environment. It hence represents an actual ``cultural commons'' for the group as a whole \citep[see][]{Hess2008}.

A final points regards the advantage of having coexisting cultures, which is mainly linked to the fact that each of them can exploit a different environmental niche.\footnote{Note that, in the real world, a dark side of cultural diversity exists, namely that interactions among different groups and cultures tend often to degenerate into conflicts \citep{RB2001,Soltis1995}.} The ``environment'' used in our model is too simple to create many niches. As a consequence, few memetic groups can be supported at the same time and instability, e.g. due to high mutation levels, quickly produces a convergence of the agent's memotypes (at least in the cultural evolution models). Nevertheless, in the \emph{Group} model two niches are, on average, present after every environment redefinition, which are promptly exploited by agents, at least under certain parameter conditions. What we found is actually a system where two different ``cultural commons'' (i.e. two different $M_g$) allow the persistence of two different memetic groups in a two-niche environment. This supports our second hypothesis, arguing that cultures coexistence is linked with the exploitation of different niches in a given social-ecological system.

The situation is somewhat more complicated in the niche construction models. A randomly defined  environment of length $N \times m=1536$ holds, on average, only 1.35 niches: a value probably too small to support the coexistence of different memetic groups. This contrasts with our finding of at least two coexisting groups for all parameter configurations in the \emph{NC-group} model. However, the environment defined in the model depends on the agents memotypes. It is hence not random and can support more niches. Any memotype in the \emph{NC-group} model creates  at least a niche for its complement. For instance, the memotype $M_1 = \{0,0,0\}$ creates a niche for the memotype $M_2 = \{1,1,1\}$ and vice-versa. The point here is that any large group creates a niche for a complementary group, building hence a two-niche environment. The observed convergence towards the coexistence of two large groups in the model should therefore not represent a surprise. It is actually the logical consequence of the niche construction process.

While more complex environments may lead to the coexistence of more groups, our model shows that the coexistence itself is possible given that (i) the environment possesses a sufficient number of niches and (ii) some process operates in order to keep the group boundaries at least partially impermeable (although it is not necessary to have perfect endogeneity as in Axelrod's model). This conditions can be checked against empirical evidence. For instance,  \citet{Rogers2008}, in a research on the evolution over time and space of the shape of Polynesian canoes, show that functional features change slower than the symbolic ones. This is consistent with our arguments. The adaptiveness of functional features depends indeed on the law of fluid dynamics and on the technological abilities of Polynesian people. Since these two factors vary little across islands, it is  likely that only one niche exists for functional features. The fact that we observe a convergence towards a single form --- carefully conserved over time and changing only as the result of occasional innovations --- it is hence consistent with our first argument. Note also that the innovations themselves, once adopted, tend to spread rapidly, leading again to a substantial convergence of forms. In absence of niches, the existing group barriers are indeed not effective to ``protect'' the diversity and the best canoe design rapidly spread from island to island fostered by the existing selection processes, e.g. differential fishing yields, migration success or simply survival of the sailors \citep[3418]{Rogers2008}.

The situation is markedly different for the symbolic traits of canoes, which presumably have little or no adaptive effect. In this case, a multitude of niches exists, all sharing the same level of adaptiveness (which is low, but this makes no difference as long as it is equal).  This reflects in their large variability, in a condition where even weak barriers are sufficient to maintain significant group differences. Note also that symbolic markers are often used by humans in order to define group boundaries \citep[221-224]{RB2005}, a fact suggesting that these features are not only the product of different memes shared by different groups, but may represent  barriers for the inter-group meme flow  as well.

While this is just an example, it shows how our model can be tested using empirical data. The crucial point is to evaluate the existence of alternative niches for each trait included in the analysis. Once that a multi-niche environment is detected, the next question becomes whether a similar number of memetic groups exists. This crucially depends on the strength of boundaries among  groups living in the environment under observation. The general prediction is that traits strictly associated with ``material'' aspects of live will vary only when real alternative niches exists, while symbolic traits are much more free to differentiate following the existing group boundaries.

\paragraph{Acknowledgements:}
The author gratefully acknowledges the comments received during the \emph{First International Workshop on Cultural Commons}, Torino, January 29--30, 2010. A special thanks goes to Lucia Tamburino for her comments and her help in developing part of the simulation codes.

\bibliographystyle{chicagoa}
\bibliography{CultEvol}

\end{document}